\documentclass[twocolumn,10pt,showpacs,notitlepage,floatfix,
superscriptaddress,amsmath,amssymb,nofootinbib,floatfix,aps,prx
]{revtex4-2}

\usepackage[normalem]{ulem}
\usepackage{mathpazo}
\usepackage{graphicx}
\usepackage{amssymb}
\usepackage{amsmath}
\usepackage{amsthm}
\usepackage{amsfonts}
\usepackage{mathtools}
\usepackage{xcolor}
\usepackage{hyperref}
\usepackage{tikz}
\usepackage{orcidlink}
\usepackage[ruled,vlined,linesnumbered]{algorithm2e}

\RestyleAlgo{ruled}
\SetKwComment{Comment}{/* }{ */}
\SetKwInput{KwRequire}{Require}
\SetKwInput{KwInput}{Input}
\SetKwInput{KwOutput}{Output}
\SetKwInput{KwVar}{Variables}
\usetikzlibrary{arrows.meta, positioning, shapes.geometric}

\usetikzlibrary{cd}

\graphicspath{{figures/}}

\newtheorem*{lem*}{Lemma}
\newtheorem*{thm*}{Theorem}

\newcommand{\eq}[1]{\hyperref[eq:#1]{Eq.~(\ref*{eq:#1})}}
\renewcommand{\sec}[1]{\hyperref[sec:#1]{Section~\ref*{sec:#1}}}
\newcommand{\secsm}[1]{\hyperref[sec:#1]{Sec.~\ref*{sec:#1}}}
\newcommand{\app}[1]{\hyperref[app:#1]{Appendix~\ref*{app:#1}}}
\newcommand{\theo}[1]{\hyperref[thm:#1]{Theorem~\ref*{thm:#1}}}
\newcommand{\algo}[1]{\hyperref[alg:#1]{Algorithm~\ref*{alg:#1}}}
\newcommand{\lemm}[1]{\hyperref[lem:#1]{Lemma~\ref*{lem:#1}}}
\newcommand{\defn}[1]{\hyperref[defn:#1]{Definition~\ref*{defn:#1}}}
\newcommand{\corr}[1]{\hyperref[cor:#1]{Corollary~\ref*{cor:#1}}}
\newcommand{\condition}[1]{\hyperref[cond:#1]{Condition~\ref*{cond:#1}}}
\newcommand{\fig}[1]{\hyperref[fig:#1]{Fig.~\ref*{fig:#1}}}
\newcommand{\tab}[1]{\hyperref[tab:#1]{Table~\ref*{tab:#1}}}
\newcommand{\tabsm}[1]{\hyperref[tab:#1]{Tab.~\ref*{tab:#1}}}
\newcommand{\propos}[1]{\hyperref[prop:#1]{Proposition~\ref*{prop:#1}}}
\newcommand{\propsm}[1]{\hyperref[prop:#1]{Prop.~\ref*{prop:#1}}}
\newcommand{\rema}[1]{\hyperref[rem:#1]{Remark~\ref*{rem:#1}}}

\newcommand{\code}{\mathcal{C}}
\newcommand{\transpose}{T}

\newcommand{\ztwo}{\mathbb{F}_2}
\newcommand{\rank}{\operatorname{rank}}

\newcommand{\image}[1]{{\rm im}\,#1}

\begin{document}

\title{Bidirectional Decoding for Concatenated Quantum Hamming Codes}

\author{Chao Zhang\orcidlink{0000-0002-2093-7496}}
\thanks{These authors contributed equally to this work.}
\affiliation{Department of Physics, The Hong Kong University of Science and Technology, Clear Water Bay, Kowloon, Hong Kong SAR, China}
\author{Zipeng Wu\orcidlink{0000-0002-9349-1325}}
\thanks{These authors contributed equally to this work.}
\affiliation{SpinQ Technology (Hong Kong) Co. Ltd., Cyberport, Pok Fu Lam, Hong Kong SAR, China}
\author{Jiahui Wu\orcidlink{0009-0009-9733-021X}}
\affiliation{Department of Physics, The Hong Kong University of Science and Technology, Clear Water Bay, Kowloon, Hong Kong SAR, China}
\author{Shilin Huang\orcidlink{0000-0001-6731-8601}}
\email{huangsl@ust.hk}
\affiliation{Department of Physics, The Hong Kong University of Science and Technology, Clear Water Bay, Kowloon, Hong Kong SAR, China}

\begin{abstract}
	High-rate concatenated quantum codes offer a promising pathway toward fault-tolerant quantum computation, yet designing efficient decoders that fully exploit their error-correction capability remains a significant challenge.
In this work, we introduce a hard-decision decoder for concatenated quantum Hamming codes with time complexity polynomial in the block length. This decoder overcomes the limitations of conventional local decoding by leveraging higher-level syndrome information to revise lower-level recovery decisions---a strategy we refer to as bidirectional decoding.
For the concatenated $[[15,7,3]]$ quantum Hamming code under independent bit-flip noise, the bidirectional decoder improves the threshold from approximately $1.56\%$ to $4.35\%$ compared with standard local decoding.
Moreover, the decoder empirically preserves the full $3^{L}$ code-distance scaling for at least three levels of concatenation, resulting in substantially faster logical-error suppression than the $2^{L+1}$ scaling offered by local decoders.
Our results can enhance the competitiveness of concatenated-code architectures for low-overhead fault-tolerant quantum computation.
\end{abstract}

\maketitle


\section{Introduction}\label{sec:introduction}

Quantum error-correcting codes~\cite{shor1995scheme,
Calderbank1996GoodQuantum,
steane1996error,
Calderbank_1997,
gottesman1997stabilizer,knill1997theory} are essential for scalable, fault-tolerant quantum computation~\cite{shor1996fault,aharonov1997fault,preskill1998reliable,gottesman1998theory}. Among existing code families, the surface code~\cite{bravyi1998quantum,kitaev2003fault,fowler2012surface} has become a leading candidate due to its compatibility with two-dimensional nearest-neighbor architectures~\cite{barends2014superconducting,zhao2022realization,google2023suppressing} and its high fault-tolerance threshold~\cite{dennis2002topological,fowler2009high,stephens2014fault}.
However, a surface code logical qubit of distance $d$ requires $O(d^2)$ physical qubits, resulting in substantial spatial overhead
~\cite{Bravyi2009,Bravyi2010,Baspin2022}. This challenge has motivated the search for quantum codes that relax geometric locality restrictions while offering higher encoding rates, such as quantum low-density-parity-check (LDPC) codes~\cite{tillich2013quantum,breuckmann2021quantum,panteleev2021quantum}.
Recent hardware advances  that support programmable nonlocal connectivity~\cite{bluvstein2024logical,sahu_entangling_2023,pino_demonstration_2021,almanakly_deterministic_2025,main_distributed_2025} make the practical realization of high-rate, nonlocal codes increasingly feasible.

Besides quantum LDPC codes, a promising pathway is the use of concatenated quantum codes~\cite{knill1996concatenated}, which historically played a central role in establishing the celebrated threshold theorem~\cite{aharonov1997fault,knill1998resilient,aliferis2005quantum}.
While early concatenated codes were often associated with low theoretical threshold bounds, Knill’s C4/C6 architecture achieved a $3\%$ threshold with all-to-all connectivity~\cite{knill2005quantum}.

Very recently, concatenated codes with high encoding rates have received increasing attention.
Yamasaki and Koashi~\cite{Yamasaki2024} demonstrated that by concatenating quantum Hamming codes with growing sizes, one can construct fault-tolerant quantum computation protocol with  constant space overhead and quasi-polylogarithmic time overhead. Another notable example is Goto’s concatenation of $[[2n,2n-2,2]]$ quantum error-detecting codes, known as many-hypercube codes~\cite{Goto2024}, which simultaneously achieve a high encoding rate and a percent-level fault-tolerance threshold.
In addition, recent work has balanced encoding efficiency and error suppression by concatenating high-threshold codes, such as surface codes, at the low level with high-rate codes at the high level~\cite{yoshida2025concatenate,gidney2025yoked,Pattison2025hierarchical,junichi2025}. While effective, this approach inevitably sacrifices the encoding rate. To realize the full potential of high-rate architectures without this overhead, it is essential to improve the decoding of the high-rate codes themselves.

However, designing decoders for high-rate concatenated codes that are both computationally efficient and accurate remains challenging.
A conventional example is the \textit{local hard-decision decoder}, which selects a single locally optimal recovery for each block using only its own syndrome.
While computationally efficient, this decoder fails to preserve the code distance by ignoring inter-block constraints imposed by higher-level syndromes. Furthermore, it treats the highly correlated logical qubits of a single block as independent inputs to the next level; this neglect of intrinsic correlations significantly lowers the threshold.

To ensure optimal decoding for concatenated codes, one should ideally defer all local recovery decisions until global syndrome information is available. Poulin's maximum-likelihood decoder achieves this by propagating the full probability distribution of errors up the concatenation hierarchy~\cite{Poulin2006}. However, tracking this complete distribution is computationally intractable for high-rate codes, as the complexity scales exponentially with the number of logical qubits.

Intermediate strategies attempt to balance this trade-off by propagating partial information, such as flagging ambiguous blocks as erasures~\cite{knill2005quantum} or tracking marginal probabilities for individual logical qubits~\cite{Goto2024}. While more efficient than maximum-likelihood decoding, these methods typically treat logical qubits as independent entities. By ignoring the strong correlations between qubits within a block, they fail to distinguish complex error patterns, resulting in thresholds that remain well below the theoretical limit.

All the decoders discussed above share a common limitation: information flows strictly \textit{one way}, from lower to higher concatenation levels. This creates a fundamental dilemma. To be efficient, a decoder must make firm decisions at lower levels, effectively discarding alternative possibilities. However, without feedback from higher levels, these early commitments cannot be revised, even if they later prove inconsistent with the global syndrome. This forces an unavoidable trade-off: one must either retain an impractical amount of information to avoid errors, or discard information to gain speed, inevitably degrading performance.

A notable exception to this one-way paradigm is Goto’s level-by-level minimum-distance decoder for many-hypercube codes~\cite{Goto2024}. This algorithm recursively constructs minimum-weight recoveries for each block. Crucially, to ensure these recoveries satisfy higher-level parity checks, the decoder does not merely combine the locally optimal choices of the subblocks. Instead, it searches for combinations where specific subblocks may adopt locally suboptimal recoveries to satisfy the syndrome, provided this minimizes the total weight. By using higher-level constraints to force revisions of lower-level decisions, the decoder effectively propagates information \textit{bidirectionally}. This approach yields high thresholds and preserves the code distance. However, it relies on enumerating valid assignments of logical corrections across subblocks. While manageable for distance-2 codes, this combinatorial search becomes intractable for codes with higher distances, limiting its applicability to general concatenated architectures.

In this work, we generalize the bidirectional decoding paradigm to the family of concatenated distance-3 quantum Hamming codes. To overcome the scalability bottleneck of prior methods, we introduce a greedy \textit{reassignment} strategy that avoids brute-force enumeration. Instead of exhaustively checking all valid logical corrections, our decoder employs a dual greedy heuristic: it greedily identifies the most promising logical adjustments to satisfy higher-level syndromes, and efficiently estimates their physical costs via a recursive greedy procedure. This approach allows the decoder to correct error patterns that local hard-decision decoding would fail, all while maintaining strictly polynomial complexity.

Our numerical results confirm that this approach yields significant performance gains. For the concatenated $[[15,7,3]]$ quantum Hamming code, we observe a threshold increase from $1.56\%$ to $4.35\%$ under independent bit-flip noise. Crucially, our decoder empirically preserves the full code distance $d=3^L$ for at least three concatenation levels, a property we verify across various heterogeneous concatenation schemes. This contrasts sharply with the reduced effective distance of $\sim 2^{L+1}$ for local decoding, resulting in much steeper error suppression. 
Consequently, at a physical bit-flip error rate of $1\%$, the bidirectional decoder applied to seven independent $3$-level concatenated $[[15,7,3]]$ code can reduce the logical error rate by five orders of magnitude compared to the local decoder applied to a single $4$-level concatenated $[[15,7,3]]$ code of equal logical capacity. Leveraging the decoder's polynomial efficiency, we extend our benchmarks to four concatenation levels ($50,625$ physical qubits), where the effective distance remains close to the theoretical limit.

The remainder of this paper is organized as follows.
Section~\ref{sec:background} reviews quantum Hamming codes and their concatenated constructions.
Section~\ref{sec:local} analyzes the limitations of the local hard-decision decoder when applied to concatenated quantum Hamming codes.
Section~\ref{sec:greedy} introduces our bidirectional hard-decision decoding algorithm.
Section~\ref{sec:results} presents numerical decoding results for concatenated quantum Hamming codes and examines the impact of heterogeneous concatenation structures.
Section~\ref{sec:discussion} discusses the underlying assumptions of the proposed decoder and the challenges associated with extending it to more general concatenated codes.
Finally, Section~\ref{sec:conclusion} concludes and outlines future research directions.

\section{Preliminaries}
\label{sec:background}

Throughout this work, all vectors and matrices are defined over the binary field
\(\mathbb{F}_2 = \{0,1\}\), with addition and multiplication performed modulo~2.
All vectors are treated as column vectors unless otherwise specified.
For a binary matrix \(H\), the transpose, kernel (null space), image (column space) and rank of $H$ are denoted by
$H^T$, $\ker H$, $\image H$ and $\rank H$ respectively.
The \emph{Hamming weight} of a binary vector \(x \in \mathbb{F}_2^n\), denoted by \(|x|\), is the number of nonzero entries in \(x\).

The symbol \(\oplus\) denotes bitwise XOR between binary vectors of the same length, or equivalently between integers written in binary representation with a fixed number of bits.
For example,
\[(011)_2 \oplus (101)_2 = (110)_2.\]
We also find it useful to define the bitwise OR operation between two binary vectors of the same length. Namely, for two vectors \(a,b \in \ztwo^n\),
\[
	(a \vee b)_i := a_i \vee b_i.
\]

\subsection{Calderbank--Shor--Steane codes}

A Calderbank--Shor--Steane (CSS) code~\cite{Calderbank1996GoodQuantum,Steane1996Multiple} acting on $n$ physical qubits is specified by two binary parity-check matrices
\(H_X \in \ztwo^{m_X \times n}\) and \(H_Z \in \ztwo^{m_Z \times n}\),
which satisfy the commutation condition
\[
	H_X H_Z^\transpose = 0 .
\]
We denote the $X$- and $Z$-stabilizer groups by $\mathcal{S}_X = \image{H_X^T}$ and $\mathcal{S}_Z = \image{H_Z^T}$.
The code parameters are denoted by $[[n,k,d]]$, where the number of encoded qubits is
\[
	k = n - \rank{H_X} - \rank{H_Z},
\]
and $d$ is the code distance. Throughout this work, we assume that both parity-check matrices have full rank,
$\rank{H_X}=m_X$ and $\rank{H_Z}=m_Z$, so that all stabilizer generators are linearly independent.

A key feature of CSS codes is that $X$-type and $Z$-type errors can be decoded independently without reducing the code distance.
Throughout this work, we restrict attention to an idealized error model in which only bit-flip ($X$-type) errors occur on the data qubits.
An $X$-type error is represented by a binary vector $e \in \ztwo^n$, with syndrome
\(
H_Z e ,
\)
and its weight is denoted by $|e|$.

The $X$-type and $Z$-type distances are defined as
\begin{align*}
	d_X & := \min\{\, |x| : x \in (\ker H_Z) \setminus \mathcal{S}_X \}, \\
	d_Z & := \min\{\, |z| : z \in (\ker H_X) \setminus \mathcal{S}_Z \},
\end{align*}
and the overall code distance is $d = \min\{d_X,d_Z\}$.

For each logical qubit $i \in [k]$, one may choose logical operators $\overline{X}_i$ and $\overline{Z}_i$, represented by binary vectors
\[
	\overline{x}_i \in (\ker H_Z) \setminus \mathcal{S}_X,
	\qquad
	\overline{z}_i \in (\ker H_X) \setminus \mathcal{S}_Z,
\]
which satisfy the canonical commutation relation
\[
	\overline{x}_i^\transpose \overline{z}_j = \delta_{ij}.
\]
Collecting these vectors defines the logical operator matrices
\[
	L_X =
	\begin{bmatrix}
		\overline{x}_1 & \cdots & \overline{x}_k
	\end{bmatrix},
	\qquad
	L_Z =
	\begin{bmatrix}
		\overline{z}_1 & \cdots & \overline{z}_k
	\end{bmatrix} .
\]
A logical $X$ operator specified by a column vector $u \in \mathbb{F}_2^k$ corresponds to the physical operator $L_X u$.
Similarly, a logical $Z$ operator $v \in \mathbb{F}_2^k$ corresponds to $L_Z v$.

For notational convenience, we write
\[
	\code := (H_X, H_Z, L_X, L_Z)
\]
to denote a CSS code together with a fixed choice of logical operators.

\subsection{Quantum Hamming codes}

Quantum Hamming codes (QHCs)~\cite{Steane1996} form a family of CSS stabilizer codes with parameters
\[
	[[n=2^r-1,\; k=2^r-2r-1,\; d=3]].
\]
The smallest member of this family is the well-known $[[7,1,3]]$ Steane code.

A convenient way to describe QHCs is to label the $n=2^r-1$ physical qubits by the nonzero $r$-bit binary strings, or equivalently by the integers $1,2,\ldots,2^r-1$. The parity-check matrices $H_X$ and $H_Z$ are identical and are constructed by taking these $r$-bit binary representations as columns. For example, the $[[7,1,3]]$ Steane code has
\begin{equation*}
	H_X = H_Z =
	\left(\begin{matrix}
			0 & 0 & 0 & 1 & 1 & 1 & 1 \\
			0 & 1 & 1 & 0 & 0 & 1 & 1 \\
			1 & 0 & 1 & 0 & 1 & 0 & 1
		\end{matrix}\right).
\end{equation*}

There is no standard choice of logical operators for QHCs. In practice, logical operator matrices $L_X$ and $L_Z$ are often obtained using Gram--Schmidt orthogonalization procedure~\cite{wilde2009logical}, which we employ in this work. Very recently, a systematic construction of a symplectic basis with $L_X = L_Z$ has been proposed~\cite{tansuwannont2025clifford}.

An important structural property of QHCs is the form of their minimum-weight logical operators. For the $[[2^r-1,\,2^r-2r-1,\,3]]$ code, any weight-$3$ logical operator of $X$ or $Z$ type acts on qubits with indices $a$, $b$, and $c$ satisfying
\[
	a \oplus b \oplus c = 0,
\]
where $\oplus$ denotes bitwise XOR on $r$-bit integers. For instance, in the Steane code,
\[
	3 \oplus 5 \oplus 6 = (011)_2 \oplus (101)_2 \oplus (110)_2 = 0,
\]
so errors on qubits $3$, $5$, and $6$ form a logical operator. The total number of such weight-$3$ logical operators of a given type is
\[
	\frac{1}{3}\binom{n}{2} = \frac{(2^r - 1)(2^r - 2)}{6} = O(n^2).
\]

This binary labeling also leads to a particularly simple decoding rule. The syndrome produced by a single $X$ error on qubit $q$ is exactly the $r$-bit binary representation of $q$, establishing a one-to-one correspondence between nonzero syndromes and single-qubit errors. As a result, hard-decision minimum-weight decoding reduces to a simple lookup operation: given a syndrome
\( s = (s_1,\ldots,s_r) \in \mathbb{F}_2^r \), the corresponding recovery acts on the qubit indexed by
\[
	q = \sum_{j=1}^r s_j\, 2^{\,r-j}.
\]
For example, in the Steane code with \( r = 3 \), the syndrome \( s = (0,1,1) \) maps to a correction on
qubit \( 2^1 + 2^0 = 3 \).

\subsection{Code concatenation}

Our primary focus in this work is on \emph{concatenated quantum Hamming codes}
(CQHCs)~\cite{Yamasaki2024}.
A CQHC with a total of $L$ concatenation levels is specified by a sequence of local
quantum Hamming codes (QHCs)
\[
	\mathcal{Q}^{(\ell)}
	:= \left(H_X^{(\ell)}, H_Z^{(\ell)}, L_X^{(\ell)}, L_Z^{(\ell)}\right),
	\qquad
	\ell \in \{1,\ldots,L\},
\]
each having parameters
\[
	[[n_\ell = 2^{r_\ell}-1,\; k_\ell = 2^{r_\ell}-2r_\ell-1,\; d_\ell = 3]].
\]

A level-$\ell$ CQHC encodes
\(
K_\ell := \prod_{\ell'=1}^{\ell} k_{\ell'}
\)
logical qubits into
\(
N_\ell := \prod_{\ell'=1}^{\ell} n_{\ell'}
\)
physical qubits and has distance
\[
	D_\ell := \prod_{\ell'=1}^{\ell} d_{\ell'} = 3^\ell.
\]
Logical qubits at level $\ell$ are labeled by elements of the index set
\[
	\mathcal{I}_\ell := [k_\ell] \times [k_{\ell-1}] \times \cdots \times [k_1],
\]
which records the hierarchical structure of the concatenation.

We now describe an explicit inductive construction that realizes these
parameters.
An instance of the level-$1$ CQHC, denoted by
$\mathcal{C}^{(1)}$, is identical to $\mathcal{Q}^{(1)}$.
Its physical qubits are referred to as \emph{level-$0$ qubits}, and its
logical qubits are referred to as \emph{level-$1$ qubits} and are labeled by
$\mathcal{I}_1 = [k_1]$.

For $\ell > 1$, an instance of the level-$\ell$ CQHC,
$\mathcal{C}^{(\ell)}$, is constructed from $n_\ell$ copies of the
level-$(\ell-1)$ concatenated code,
\[
	\mathcal{C}^{(\ell-1)}_{1},\ldots,\mathcal{C}^{(\ell-1)}_{n_\ell}.
\]
Each copy $\mathcal{C}^{(\ell-1)}_{i}$ contains $K_{\ell-1}$ logical qubits
labeled by $\mathcal{I}_{\ell-1}$.
Consequently, the level-$\ell$ code contains $n_\ell K_{\ell-1}$ level-$(\ell-1)$
qubits, which can be indexed by pairs $(i,\lambda)$, where
$i \in [n_\ell]$ labels the copy and $\lambda \in \mathcal{I}_{\ell-1}$ labels the logical
qubit within that copy.

For each fixed $\lambda \in \mathcal{I}_{\ell-1}$, the collection
\[
	\{(i,\lambda) : i \in [n_\ell]\}
\]
forms the physical qubits of a level-$\ell$ local QHC block
$\mathcal{Q}^{(\ell)}_{\lambda}$, which encodes $k_\ell$ logical qubits corresponding to
the level-$\ell$ qubits labeled by
\[
	\mathcal{I}_\ell = [k_\ell] \times \mathcal{I}_{\ell-1}.
\]

For an $L$-level CQHC, it is convenient to adopt a unified labeling convention for
qubits across all concatenation levels.
We use the notation
\[
	(i_L,\ldots,i_1)_\ell
\]
to denote a level-$\ell$ qubit of the level-$L$ code.
The indices $(i_L,\ldots,i_{\ell+1})$ encode the hierarchical containment of this
qubit within the concatenation tree: the level-$(L-1)$ block
$\mathcal{C}^{(L-1)}_{i_L}$ is the $i_L$-th subblock of the top-level block
$\mathcal{C}^{(L)}$; more generally, for each $\ell' \in \{\ell+1,\ldots,L\}$, the block
$\mathcal{C}^{(\ell'-1)}_{i_L,\ldots,i_{\ell'}}$ is the $i_{\ell'}$-th subblock of
its parent block $\mathcal{C}^{(\ell')}_{i_L,\ldots,i_{\ell'+1}}$.
The remaining indices $(i_\ell,\ldots,i_1) \in \mathcal{I}_{\ell}$ label the logical qubit within the specified
level-$\ell$ subblock.

\section{Local hard-decision decoding and its failure mechanism}\label{sec:local}

\begin{figure*}[ht!]
	\centering
	\includegraphics[width=\linewidth]{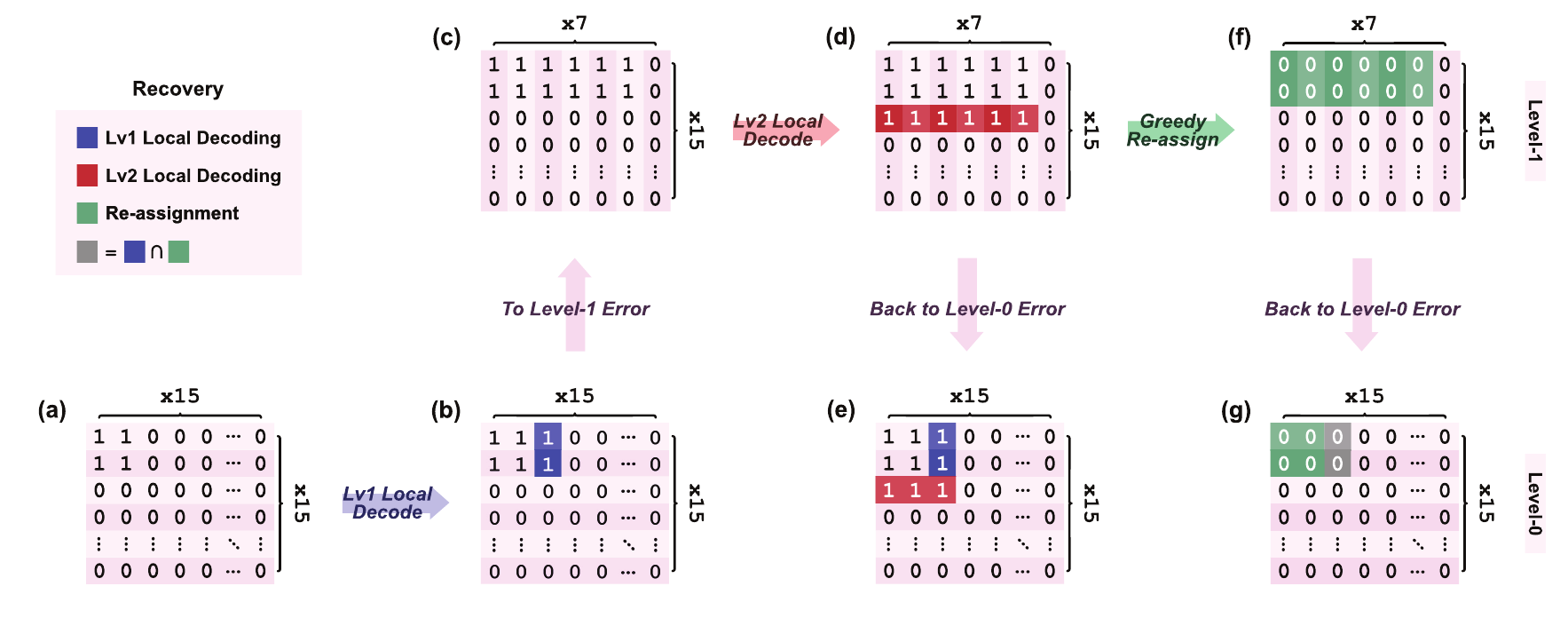}
	\caption{\label{fig:decode_failure}
		Decoding of a weight-$4$ error in a two-level concatenation of the \([[15,7,3]]\) quantum Hamming code.
		The level-$0$ qubits form a $15 \times 15$ array, while the level-$1$ qubits form a $15 \times 7$ array.
		(a) The four physical errors occur on the level-$0$ qubits $(1,1)_0$, $(1,2)_0$, $(2,1)_0$, and $(2,2)_0$.
		(b) Level-$1$ local decoding applies recoveries on the level-$0$ qubits $(1,3)_0$ and $(2,3)_0$ (shown in blue).
		(c) The resulting level-$1$ errors are concentrated on the first two rows (subblocks) and share the same pattern.
		(d) Level-$2$ local decoding propagates this error pattern to the third level-$1$ subblock (shown in red).
		(e) At the physical level, level-$2$ local decoding introduces an additional weight-$3$ recovery (shown in red), resulting in a total recovery weight of $5$.
		(f) By reassigning the level-$1$ recoveries on the first and second subblocks in a manner consistent with the level-$1$ syndrome, the error can be corrected.
		(g) The reassigned recovery has physical weight $6 - 2 = 4$.
		This example illustrates the core mechanism of the bidirectional decoder: by using higher-level syndrome information to revise lower-level decisions (reassignment), it successfully corrects error patterns that cause the local hard-decision decoder to fail.
	}
\end{figure*}

Throughout this work, \emph{local hard-decision decoding} refers to the conventional
decoding strategy for concatenated codes:
At the lowest concatenation level, a minimum-weight error (MWE) decoder is
applied independently to each level-$1$ local QHC block.
Logical failures of some level-$1$ blocks manifest as effective physical errors on
the level-$2$ local QHC blocks.
The decoder then applies the same MWE decoding procedure independently to each level-$2$
local QHC block and proceeds to higher concatenation levels.
Crucially, the decoder at level $\ell$ has no access to lower-level syndromes or
recovery choices.

Local hard-decision decoding is computationally efficient and naturally parallelizable.
However, this approach does not in general preserve the full distance of a concatenated code.
A $2$-level CQHC is sufficient to illustrate this limitation: an optimal decoder for this code can correct any error of weight up to $4$, whereas the local hard-decision decoder may fail on certain weight-$4$ error patterns.

Consider the following weight-$4$ error acting on the set of level-$0$ qubits
\[
	\{(1,1)_{0}, (1,2)_{0}, (2,1)_{0}, (2,2)_{0}\}\]
shown in Fig.~\ref{fig:decode_failure}(a).
That is, the first and second level-$1$ blocks, $\mathcal{C}_1^{(1)}$ and
$\mathcal{C}_2^{(1)}$, each contain two bit-flip errors on their first and second
physical qubits.
For a QHC, two errors on qubits $1$ and $2$ produce a syndrome corresponding to a
single-qubit error on qubit $3$, since $1 \oplus 2 = 3$.
As a result, the local MWE decoders for $\mathcal{C}_1^{(1)}$ and
$\mathcal{C}_2^{(1)}$ apply recoveries that flip the level-$0$ qubits $(1,3)_{0}$
and $(2,3)_{0}$, respectively, as shown in Fig.~\ref{fig:decode_failure}(b).
After level-$1$ local decoding, the residual error therefore acts on
\[
	\{(1,1)_{0}, (1,2)_{0}, (1,3)_{0},\;
	(2,1)_{0}, (2,2)_{0}, (2,3)_{0}\}.
\]
This residual error induces identical logical error patterns on the two affected
level-$1$ blocks, illustrated in Fig.~\ref{fig:decode_failure}(c).
The pattern can be represented by a binary vector
$v \in \mathbb{F}_2^{k_1}$.
Equivalently, the errors are supported on the level-$1$ qubits
\[
	\{(i,j)_{1} : v_j = 1,\; i \in \{1,2\},\; j \in [k_1]\}.
\]

For each level-$2$ local QHC block $\mathcal{Q}_j^{(2)}$ with $v_j = 1$, errors are
present on its qubits, namely the level-$1$ qubits $(1,j)_{1}$ and
$(2,j)_{1}$.
The level-$2$ local decoder therefore applies a recovery on the level-$1$ qubit
$(3,j)_{1}$.
As shown in Fig.~\ref{fig:decode_failure}(d),
the logical $X$ values
of the level-$1$ block
$\mathcal{C}_3^{(1)}$ is \textit{flipped} by the same pattern $v$, whose physical implementation is supported on the $3$ level-$0$ qubits
\[
	\{(3,j)_{0} : j \in \{1,2,3\}\},
\]
as shown in Fig.~\ref{fig:decode_failure}(e).
As a result, the $2$-level code $\mathcal{C}^{(2)}$ acquires a logical error acting on
the level-$2$ qubits
\[
	\{(i,j)_{2} : v_i = 1,\; v_j = 1\},
\]
whose minimum-weight physical realization is supported on $9$ level-$0$ qubits
\[
	\{(i,j)_{0} : i,j \in \{1,2,3\}\}.
\]
We can see that local hard-decision decoding applies a suboptimal physical recovery of weight $9-4=5$,
whereas an optimal decoder should correctly identify and correct the
original weight-$4$ error pattern.

More generally, any weight-$2^{L}$ bit-flip error supported on level-$0$ qubits of the form
\[
	\{(i_L,\ldots,i_1)_0 : i_\ell \in \{a_\ell,b_\ell\}\},
\]
with $a_\ell \neq b_\ell$ for all $\ell$, inevitably leads to a logical failure under local hard-decision decoding.
Consequently, although the concatenated code has distance $3^{L}$, the effective distance achieved by the local hard-decision decoder is reduced to $2^{L+1}-1$.

\section{Bidirectional hard-decision decoding}\label{sec:greedy}

The weight-$4$ error pattern
\[
	\{(1,1)_{0}, (1,2)_{0}, (2,1)_{0}, (2,2)_{0}\}
\]
in the two-level CQHC discussed in Sec.~\ref{sec:local} illustrates a key point:
the weight of a recovery evaluated at an intermediate concatenation level does
not reliably reflect its actual cost at the physical level.

After the level-$1$ local QHC decoder applies recoveries to the level-$0$ qubits $(1,3)_0$ and $(2,3)_0$, the resulting syndrome at level~$2$ generally admits multiple possible recovery configurations on level-$1$ qubits.
As discussed above, the straightforward choice adopted by the local
hard-decision decoder is to flip the logical value of the level-$1$ block
$\mathcal{C}_3^{(1)}$.
A minimum-weight physical realization of this \emph{logical flip} acts on the
level-$0$ qubits
\[
	\{(3,1)_{0}, (3,2)_{0}, (3,3)_{0}\}
\]
and has weight~3.
When combined with the existing recoveries on the level-$0$ qubits
$(1,3)_{0}$ and $(2,3)_{0}$, the total physical recovery has weight~5.

As shown in Fig.~\ref{fig:decode_failure}(f), an alternative choice is to apply the same logical flip to the level-$1$ blocks
$\mathcal{C}^{(1)}_1$ and $\mathcal{C}^{(1)}_2$ instead.
When this logical flip is described at level-$1$, it is not immediately favored, as it appears to double the recovery weight.
However, the corresponding physical recovery operator has support on
\[
	\{(1,1)_{0}, (1,2)_{0}, (1,3)_{0},\;
	(2,1)_{0}, (2,2)_{0}, (2,3)_{0}\},
\]
which overlaps with the existing recoveries on $(1,3)_{0}$ and
$(2,3)_{0}$ and therefore cancels them, as shown in Fig.~\ref{fig:decode_failure}(g).
As a result, the net physical recovery matches the
original error and has weight $4$.

The comparison above shows that identifying the correct recovery requires
augmenting local hard-decision decoding with two additional algorithmic
ingredients:
\begin{itemize}
	\item[(i)] For a prescribed logical flip $\Delta$ on a CQHC block,
	      evaluate the minimum physical recovery weight associated with realizing $\Delta$
	      in combination with the block’s existing recovery,
	      properly accounting for possible cancellations.
	\item[(ii)] When decoding a level-$\ell$ CQHC block, use the physical
	      recovery cost evaluated in (i) as a criterion to \textit{reassign} logical flips
	      among its level-$(\ell-1)$ subblocks, selecting a syndrome-consistent configuration
	      that minimizes the total physical recovery weight.
\end{itemize}

\begin{figure}[t]
	\centering
	\includegraphics[width=\linewidth]{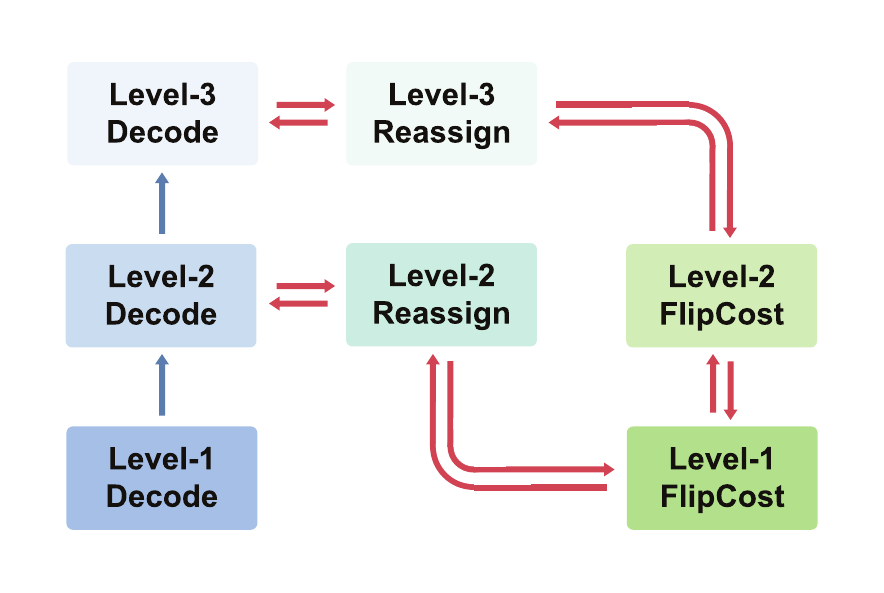}
	\caption{\label{fig:decode_birect} The information flow in bidirectional hard-decision decoding for a 3-level concatenated quantum Hamming code. Unlike standard local decoding which flows strictly bottom-up (blue arrows), the bidirectional decoder incorporates a top-down "reassignment" pass (red arrows). This allows the decoder to update lower-level recovery choices based on global constraints, significantly improving error correction performance.}
\end{figure}

Our proposed decoder implements the two algorithmic ingredients described above.
It is organized around a recursive main procedure,
$\mathsf{Decode}(\code^{(\ell)})$, which takes as input a level-$\ell$ CQHC block
together with its measured syndrome.
The pseudocode for \textsf{Decode} is shown in
Procedure~\ref{proc:decode}.

At the lowest concatenation level ($\ell=1$), the decoder computes a minimum-weight physical recovery, stores it in $R[\code^{(\ell)}]$, and terminates.
For $\ell>1$, the decoder first invokes \textsf{Decode} recursively on each level-$(\ell-1)$ subblock $\code^{(\ell-1)}_{i}$ to determine their recoveries.
These subblock recoveries induce a syndrome for the level-$\ell$ code.
Recall that the level-$\ell$ code is composed of independent local QHCs $\mathcal{Q}^{(\ell)}_{\lambda}$, one for each logical qubit $\lambda \in \mathcal{I}_{\ell-1}$ of the subblocks.
The decoder applies a Minimum Weight Error (MWE) correction to each local code $\mathcal{Q}^{(\ell)}_{\lambda}$ independently.
The resulting corrections form the columns of the recovery tensor $R[\code^{(\ell)}]$.
Specifically, the entry $R[\code^{(\ell)}]_{i,\lambda}=1$ if and only if the level-$(\ell-1)$ qubit $(i,\lambda)$ is corrected.
The $i$-th slice $R[\code^{(\ell)}]_{i,\bullet}$ then specifies the logical flip assigned to the level-$(\ell-1)$ subblock $\code^{(\ell-1)}_{i}$.

The subroutine $\mathsf{Reassign}(\code^{(\ell)})$ is then invoked to
redistribute these logical flips among level-$(\ell-1)$ subblocks using a
greedy procedure, thereby implementing ingredient~(ii).
To evaluate the physical cost of candidate reassignments,
$\mathsf{Reassign}$ calls the subroutine $\mathsf{FlipCost}$, which greedily evaluates the minimum physical recovery weight
compatible with a prescribed logical flip on a code block and thus implements (i).
After the recovery tensor $R[\code^{(\ell)}]$ has been finalized,  level-$(\ell-1)$ qubits $(i,\lambda)$ for which
$R[\code^{(\ell)}]_{i,\lambda}=1$ are corrected.

The local hard-decision decoder propagates information strictly in one direction, from lower to higher concatenation levels.
By contrast, the reassignment of logical flips in our decoding algorithm enables an on-demand flow of information from higher to lower levels, allowing higher-level constraints to revise earlier local decisions.
This bidirectional exchange substantially reduces the amount of information that must be propagated, in comparison with Poulin’s maximum-likelihood decoder.
For this reason, we refer to the proposed decoder as a \emph{bidirectional} hard-decision decoder, as illustrated schematically in Fig.~\ref{fig:decode_birect}.

We now provide the detailed description of the subroutines \(\mathsf{Reassign}\) and \(\mathsf{FlipCost}\) in Sec.~\ref{sec:main_proc} and Sec.~\ref{sec:shift}, respectively.

\begin{algorithm}[t]
	\caption{\label{proc:decode}$\mathsf{Decode}(\code^{(\ell)})$}

	\If{$\ell = 1$}
	{
		$R[\code^{(\ell)}] \gets \mathsf{MWE}(\code^{(\ell)})$\;
		\textbf{return}
	}
	\ForEach{$i = 1,\ldots,n_\ell$}{
		$\mathsf{Decode}(\code^{(\ell-1)}_{i})$
	}

	\ForEach{$\lambda \in \mathcal{I}_{\ell-1}$}{
		$R[\code^{(\ell)}]_{\bullet,\lambda} \gets \mathsf{MWE}\left(\mathcal{Q}^{(\ell)}_{\lambda} \right)$
	}

	$\mathsf{Reassign}\left(\code^{(\ell)}\right)$
\end{algorithm}

\subsection{Subroutine $\mathsf{Reassign}$}\label{sec:main_proc}

The subroutine $\mathsf{Reassign}$ seeks to modify the assignment of logical flips to level-$(\ell-1)$ subblocks in a way that preserves the level-$\ell$ syndrome but minimizes the total physical recovery weight.
The pseudocode of $\mathsf{Reassign}$ is shown in Procedure~\ref{proc:reassign}; we now explain its operation in detail.

The procedure begins by calling the subroutine
$\mathsf{FlipCost}\!\left(\code^{(\ell-1)}_i,\, R[\code^{(\ell)}]_{i,\bullet}\right)$
to evaluate the physical cost $w_i$ associated with the logical flip
$R[\code^{(\ell)}]_{i,\bullet}$ when combined with the existing recovery on
$\code^{(\ell-1)}_{i}$, explicitly accounting for possible cancellations.
The detailed description of $\mathsf{FlipCost}$ is deferred to
Sec.~\ref{sec:shift}.

The algorithm then greedily reassigns logical flips by utilizing the weight-3 logical operators of the level-$\ell$ local QHC. It iterates over all triples $(a,b,c)$ satisfying $a \oplus b = c$, which correspond to the support of such operators.
For any such triple, applying a logical flip to subblock $\code^{(\ell-1)}_{c}$ yields the same level-$\ell$ syndrome as applying the same flip to both $\code^{(\ell-1)}_{a}$ and $\code^{(\ell-1)}_{b}$.
Since this structural relationship holds for all $\lambda \in \mathcal{I}_{\ell-1}$ simultaneously, the decoder can transfer the entire logical flip vector $R[\code^{(\ell)}]_{c,\bullet}$ at once.
This property allows the decoder to \textit{transfer} a logical flip from $c$ to the pair $\{a,b\}$ while maintaining syndrome consistency.
For each triple, the algorithm invokes \textsf{FlipCost} to compare the physical cost of the current configuration against the transferred one. The transfer is accepted if and only if the total physical cost strictly decreases.
The algorithm re-iterates over all eligible triples whenever a transfer is accepted, and terminates only when no further transfer occurs.

\begin{algorithm}[t]
	\caption{\label{proc:reassign}$\mathsf{Reassign}(\code^{(\ell)})$}

	\ForEach{$i = 1, \ldots, n_\ell$}{
		$w_i \gets
			\mathsf{FlipCost}\left(\code^{(\ell-1)}_{i}, R[\code^{(\ell)}]_{i,\bullet}\right)$
	}

	\Repeat{\textnormal{$R[\code^{(\ell)}]$ has no update}}{

		\ForEach{$a\oplus b=c$ \textnormal{\bf with} $(T := R[\code^{(\ell)}]_{c,\bullet})\ne 0$}{%
			\ForEach{$i \in \{a,b,c\}$}{
				$w_i' \gets
					\mathsf{FlipCost}\left(\code^{(\ell-1)}_{i},
					R[\code^{(\ell)}]_{i,\bullet} + T\right)$
			}

			\If{$\sum_{i \in \{a,b,c\}} w_i'
					< \sum_{i \in \{a,b,c\}} w_i$}{%

				\ForEach{$i \in \{a,b,c\}$}{

					$w_i \gets w_i'$\;
					$R[\code^{(\ell)}]_{i,\bullet} \gets R[\code^{(\ell)}]_{i,\bullet} + T$
				}

				\textbf{break}
			}
		}
	}
\end{algorithm}

\subsection{Subroutine $\mathsf{FlipCost}$}\label{sec:shift}

To make reassignment valid, we must address the following task.
Given an $[[n,k]]$ CSS code $\code=(H_X,H_Z,L_X,L_Z)$, a prescribed
logical flip represented by a column vector $\Delta\in\ztwo^{k}$, and an existing
recovery $R[\code]\in\ztwo^{n}$ on the physical qubits, we seek a
syndrome-equivalent physical realization of the shift whose combination with
$R[\code]$ has minimum physical weight.
Any such realization can be written as
\[
	R[\code] + L_X \Delta + h,
\]
where $h\in\mathcal{S}_X = \image{H_X^T}$ is an $X$-stabilizer element of $\code$.
Accordingly, the quantity of interest is the coset minimum
\[
	\min_{h\in\mathcal{S}_X}
	\bigl|R[\code] + L_X \Delta + h\bigr|.
\]

For large concatenated code blocks, the stabilizer group grows exponentially
with block size, so an exhaustive search over $h\in\mathcal{S}_X$ is computationally
infeasible.
To address this challenge, we develop a heuristic algorithm that exploits the
concatenated structure of the code.
The key observation is that a low-weight physical realization of a prescribed
logical flip $\Delta$, when combined with the existing recovery, should induce
nontrivial logical flips on as few level-$(\ell-1)$ subblocks as possible.
This objective can be evaluated entirely using the representation on
level-$(\ell-1)$ qubits. The resulting heuristic is summarized in
Procedure~\ref{proc:flipcost}, which we now explain step by step.

At $\ell=1$, the subroutine directly enumerates all
$X$-stabilizer elements of $\code^{(1)} = \mathcal{Q}^{(1)}$ and returns
\[
	\min_{h \in \mathcal{S}_X^{(1)}}
	\left| R[\code^{(1)}] + L_X^{(1)} \Delta + h \right|.
\]

For $\ell>1$, the logical flip
$\Delta \in \ztwo^{k_\ell \times \cdots \times k_1}$ is an order-$\ell$ binary
tensor.
Since the level-$\ell$ code is constructed from local QHCs acting on level-$(\ell-1)$ logical qubits, it is convenient to view $\Delta$ as a collection of column vectors $\Delta_{\bullet,\lambda} \in \ztwo^{k_\ell}$, indexed by $\lambda \in \mathcal{I}_{\ell-1}$.
The subroutine first represents the effect of $\Delta$, together with the
existing recovery $R[\code^{(\ell)}]$, on these level-$(\ell-1)$ qubits.
Specifically, for each index $\lambda \in \mathcal{I}_{\ell-1}$ labeling a logical qubit within the subblocks, the algorithm constructs a column vector
\[
	F_{\bullet,\lambda}
	= L_X^{(\ell)} \Delta_{\bullet,\lambda}
	+
	R[\code^{(\ell)}]_{\bullet,\lambda}.
\]
$F_{i,\lambda}$ records the logical flip induced on
level-$(\ell-1)$ qubit $(i,\lambda)$ by the prescribed $\Delta$ and the existing recovery.

The objective is to modify each column $F_{\bullet,\lambda}$ by adding an
$X$-stabilizer element $h_{\lambda} \in \mathcal{S}_X^{(\ell)}$ of the level-$\ell$ local QHC $\mathcal{Q}_\lambda^{(\ell)}$ so as to minimize the number of level-$(\ell-1)$ subblocks $\code^{(\ell-1)}_{i}$ whose logical value needs to be flipped.
Finding the globally optimal choice of stabilizers is yet computationally
infeasible. Therefore, the subroutine employs a greedy heuristic described
below.

\begin{algorithm}[t]
	\caption{\label{proc:flipcost}$\mathsf{FlipCost}(\code^{(\ell)},\Delta)$}

	\If{$\ell = 1$}{
	$\textbf{return} \min_{h \in \mathcal{S}_{X}^{(1)}}
		\left|R[\mathcal{C}^{(\ell)}] + L_{X}^{(1)} \Delta + h \right|$\;
	}

	\ForEach{$\lambda \in \mathcal{I}_{\ell-1}$}{
		$F_{\bullet,\lambda}
			\gets L_{X}^{(\ell)}\Delta_{\bullet,\lambda}
			+ R[\code^{(\ell)}]_{\bullet,\lambda}$\;
	}
	Fix a permutation $\alpha$ of $\mathcal{I}_{\ell-1}$ such that
	\[
		\left|R[\code^{(\ell)}]_{\bullet,\alpha_1}\right|
		\ge \cdots \ge
		\left|R[\code^{(\ell)}]_{\bullet,\alpha_{\left|\mathcal{I}_{\ell-1}\right|}}\right|;
	\]

	$F^{\mathrm{opt}} \gets F$\;

	\ForEach{$h_{\alpha_1} \in \mathcal{S}_{X}^{(\ell)}$}{
	$F'_{\bullet,\alpha_1} \gets F_{\bullet,\alpha_1} + h_{\alpha_1}$\;
	$m \gets F'_{\bullet,\alpha_1}$\;
	\ForEach{$j = 2,\ldots,\left|\mathcal{I}_{\ell-1}\right|$}{
	$h_{\alpha_j} \gets
		\arg\min_{h \in \mathcal{S}_{X}^{(\ell)}}
		\left|m \vee \left( F_{\bullet,\alpha_j} + h\right)\right|$\;
	$F'_{\bullet,\alpha_j} \gets F_{\bullet,\alpha_j} + h_{\alpha_j}$\;
	$m \gets m \vee F'_{\bullet,\alpha_j}$
	}

	\If{$\left|s\right|
			< \left|\bigvee_{\lambda \in \mathcal{I}_{\ell-1}}
			F^{\mathrm{opt}}_{\bullet,\lambda}\right|$}{
		$F^{\mathrm{opt}} \gets F'$\;
	}
	}

	\textbf{return} $ \sum_{i=1}^{n_\ell} \mathsf{FlipCost}\!\left(
		\mathcal{C}^{(\ell-1)}_{i},
		F_{i,\bullet}^{\mathrm{opt}}
		\right)$
\end{algorithm}

To implement this minimization, the heuristic fixes the stabilizer choices
column by column.
The columns of $F$ are processed in an order
$\alpha = (\alpha_1,\ldots,\alpha_{|\mathcal{I}_{\ell-1}|})$, which is a
permutation of the index set $\mathcal{I}_{\ell-1}$.
We find it effective to choose $\alpha$ such that
\[
	\left| R[\code^{(\ell)}]_{\bullet,\alpha_1} \right|
	\ge \cdots \ge
	\left| R[\code^{(\ell)}]_{\bullet,\alpha_{|\mathcal{I}_{\ell-1}|}} \right|,
\]
so that columns with larger initial weight are processed first.

Suppose that stabilizer elements have already been fixed for the first
$j-1$ columns.
When processing column $\alpha_j$, the algorithm considers all possible choices
of $X$-stabilizer elements $h_{\alpha_j} \in \mathcal{S}_X^{(\ell)}$ of the level-$\ell$ local QHC code and
temporarily updates the column according to
\[
	F_{\bullet,\alpha_j} \;\leftarrow\;
	F_{\bullet,\alpha_j} + h_{\alpha_j},
\]
while keeping the previously fixed columns unchanged.
For each candidate choice, the algorithm evaluates the cumulative row-wise
logical support of the columns processed so far via the row-wise logical OR
\[
	m = \bigvee_{j'=1}^{j} F_{\bullet,\alpha_{j'}}.
\]
The Hamming weight $|m|$ equals the number of level-$(\ell-1)$ subblocks that
undergo a nontrivial logical flip.
The greedy rule selects the stabilizer element $h_{\alpha_j}$ that minimizes
this weight.
Once selected, the choice is fixed and the procedure proceeds to the next
column.

In practice, immediately fixing the stabilizer element associated with the
first column according to the greedy rule can lead to local minima.
To mitigate this issue, the algorithm introduces a one-step lookahead at the
first column.
Rather than fixing the stabilizer choice for the first column outright, the
algorithm iterates over all possible stabilizer elements associated with that
column.
For each candidate choice, the greedy procedure is then applied to the
remaining columns, and the resulting total number of affected level-$(\ell-1)$
subblocks is evaluated.
Among all candidates, the resulting configuration $F^{\mathrm{opt}}$ whose
associated greedy solution minimizes this number is selected.

Once $F^{\mathrm{opt}}$ has been determined, each slice
$F^{\mathrm{opt}}_{i,\bullet}$ specifies the logical flip to be applied to the
level-$(\ell-1)$ subblock $\code^{(\ell-1)}_i$.
The subroutine then evaluates the total physical recovery cost by recursively
calling
\[
	\mathsf{FlipCost}\!\left(
	\code^{(\ell-1)}_i,\; F^{\mathrm{opt}}_{i,\bullet}
	\right),
	\qquad i=1,\ldots,n_\ell,
\]
and summing the resulting values.
The returned sum is the physical recovery weight associated with the prescribed
logical flip $\Delta$ on $\code^{(\ell)}$.

\subsection{Complexity analysis}\label{sec:complexity}

To analyze the time complexity of the bidirectional decoder, we proceed hierarchically.
We first analyze the subroutine $\mathsf{FlipCost}$, then the subroutine
$\mathsf{Reassign}$, and finally the overall decoding procedure.
To simplify our analysis, we assume that all concatenation levels use the same $[[n = 2^r-1,k=2^r-2r-1]]$ local QHC code.

\subsubsection{Complexity of $\mathsf{FlipCost}$}

A key structural property is that the $X$-stabilizer group of the local QHC has cardinality
\[
	\left|\mathcal{S}_X\right|
	= 2^{r}
	= n + 1
	= O(n).
\]
We also recall that the number of level-$(\ell-1)$ logical qubits is
\[
	K_{\ell-1}
	= |\mathcal{I}_{\ell-1}|
	= k^{\ell-1}.
\]

For $\ell>1$, the time complexity of $\mathsf{FlipCost}$ is dominated by the greedy
heuristic.
Fix an initial choice $h_{\alpha_1} \in \mathcal{S}_X^{(\ell)}$.
For each subsequent column $\alpha_j$ with $j \ge 2$, the algorithm computes
\[
	h_{\alpha_j} \in
	\arg\min_{h\in\mathcal{S}_X^{(\ell)}}
	\Bigl|\, m \vee (F_{\bullet,\alpha_j}+h)\Bigr|.
\]
Evaluating the objective function for a single candidate $h$ requires forming
$F_{\bullet,\alpha_j}+h$, computing its bitwise OR with $m$, and evaluating the
Hamming weight.
All of these operations act on length-$n$ binary vectors and therefore require
$O(n)$ time.

Enumerating all candidates involves
$\left|\mathcal{S}_X^{(\ell)}\right| = O(n)$ stabilizers.
Consequently, a single $\arg\min$ computation costs
\[
	O\!\left(\left|\mathcal{S}_X^{(\ell)}\right| \cdot n\right)
	=
	O(n^2).
\]
This procedure is repeated for $K_{\ell-1}-1 = O(K_{\ell-1})$ columns.
Thus, the greedy completion following a fixed initial choice of $h_{\alpha_1}$
incurs a cost
\[
	O(k^{\ell-1} n^2).
\]

The one-step lookahead strategy enumerates all possible choices of $h_{\alpha_1}$
and executes the greedy completion for each.
Therefore, the total cost of the greedy heuristic at level $\ell$ is
\begin{equation*}
	\label{eq:greedy_level_cost}
	O\!\left(\left|\mathcal{S}_X^{(\ell)}\right|
	\cdot k^{\ell-1} n^2\right)
	=
	O(k^{\ell-1} n^3).
\end{equation*}

After obtaining the optimized matrix $F^{\mathrm{opt}}$, the subroutine returns
\[
	\sum_{i=1}^{n_\ell}
	\mathsf{FlipCost}\!\left(\code^{(\ell-1)}_i,\,
	F^{\mathrm{opt}}_{i,\bullet}\right),
\]
which induces the recurrence relation
\[
	T_\ell
	=
	O(K_{\ell-1} n^3)
	+ n T_{\ell-1}.
\]
The base case is
\[
	T_1
	=
	O\!\left(|\mathcal{S}_X^{(1)}|\, n\right)
	=
	O(n^2),
\]
which follows from direct enumeration of
$|\mathcal{S}_X^{(1)}| = n+1$ stabilizers, together with $O(n)$ work per candidate
for binary vector operations.

Solving the recurrence yields
\begin{eqnarray*}
	T_\ell
	&=&
	\sum_{\ell'=1}^{\ell}
	n^{\ell-\ell'}
	O(k^{\ell'-1} n^3)\\
	&=&
	O\left( n^{\ell+2}\sum_{\ell'=1}^{\ell} \left(\frac{k}{n}\right)^{\ell'-1}\right) = O\left(n^{\ell+2}\right)
\end{eqnarray*}
as the geometric series $\sum_{\ell'=1}^{\ell} (k/n)^{\ell'-1}$ is bounded by a constant (since $k < n$).

\subsubsection{Complexity of $\mathsf{Reassign}$}

When $\mathsf{Reassign}$ is invoked, each iteration of the
subroutine calls $\mathsf{FlipCost}$ at most $O(n^2)$ times.
Moreover, in each iteration the total weight of the physical recovery is
guaranteed to decrease by at least one.
A very loose upper bound on the number of iterations is therefore
$O(N_\ell) = O(n^\ell)$, representing the worst-case scenario where the decoder corrects the entire code block one flip at a time.
Combining these estimates yields the following loose upper bound on the time
complexity of $\mathsf{Reassign}$ at level~$\ell$:
\[
	O\!\left(n^2 \cdot n^\ell\right) \cdot T_\ell
	=
	O\!\left(n^{2\ell+4}\right).
\]

\subsubsection{Complexity of $\mathsf{Decode}$}
To obtain a worst-case complexity bound for the main decoding procedure, we
assume that the cost of each invocation of $\mathsf{Decode}$ is dominated by
the reassignment subroutine.
Since a level-$\ell$ instance of $\mathsf{Decode}$ is invoked only once for
each level-$\ell$ subblock of the level-$L$ code, the total time complexity is
bounded by
\[
	\sum_{\ell=1}^L
	n^{L-\ell}
	\, O\!\left(n^{2\ell+4}\right)
	=
	O\!\left(n^{2L+4}\right).
\]
An $L$-level CQHC has block size $N = n^L$.
Therefore, the worst-case time complexity scales as
\(
	O\!\left(N^2 n^4\right)
\),
which is polynomial in the block size.

\subsubsection{Expected time complexity analysis}

In practice, the reassignment procedure is triggered only when the level-$\ell$
local codes yield nontrivial syndromes.
This occurs only if at least one of the level-$(\ell-1)$ subblocks has incurred
a logical error.
Assuming that the physical error rate $p$ lies well below the threshold
$p_{\mathrm{th}}$, we adopt the standard ansatz for concatenated codes that the
logical error rate at level~$\ell$ scales as
\[
	p_{\mathrm{L}}^{(\ell)} \propto \Lambda^{3^\ell},
	\qquad
	\Lambda := \left(p / p_{\mathrm{th}}\right)^{1/2}.
\]
The reassignment procedure at level $\ell$ is triggered if any of the $n$ level-$(\ell-1)$ subblocks fails. By the union bound, this probability scales as $n p_{\mathrm{L}}^{(\ell-1)} \propto n \Lambda^{3^{\ell-1}}$.
Under this assumption, the expected time complexity is bounded by
\[
	\sum_{\ell=1}^L
	n^{L-\ell}
	\, O\!\left(n^{2\ell+4}\right)
	\cdot n \Lambda^{3^{\ell-1}}
	=
	O\!\left(
	N\sum_{\ell=1}^L
	n^{\ell+5}\,
	\Lambda^{3^{\ell-1}}
	\right).
\]
When $\Lambda \ll 1$, the factor $\Lambda^{3^{\ell-1}}$ decays superexponentially
with $\ell$, overpowering the geometric growth of the polynomial factors in $n$.
Consequently, the sum
\[
\sum_{\ell=1}^L n^{\ell+5}\,\Lambda^{3^{\ell-1}}
\]
is bounded by \(O(n^{\beta})\) for some constant \(\beta\) that depends on \(\Lambda\) but is independent of the total number of levels \(L\).
The total expected time complexity of the bidirectional decoder is therefore
\(O(N n^{\beta}) = o(N^2)\). 

\begin{figure*}[t]
	\centering
	\includegraphics[width=\linewidth]{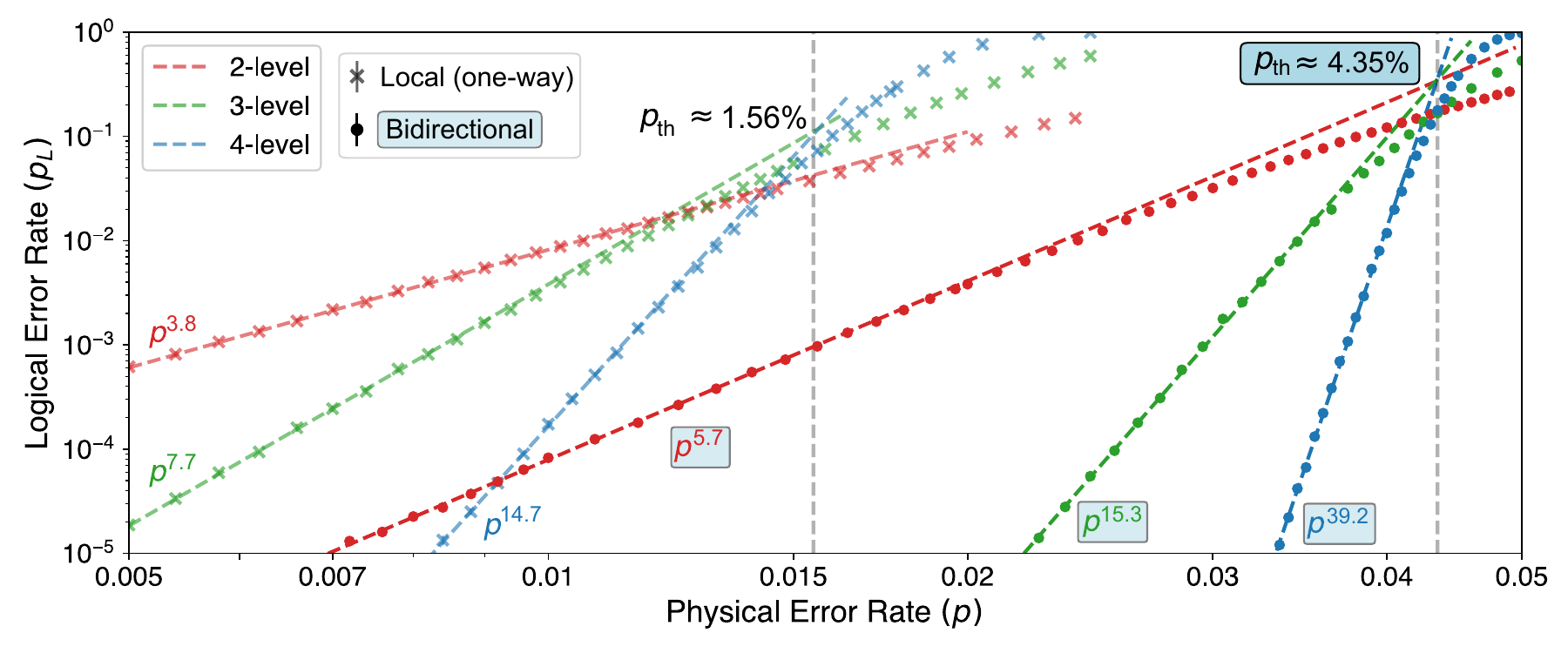}
	\caption{\label{fig:hamming1a2}Performance comparison between the local hard-decision decoder (cross markers) and the bidirectional decoder (dot markers) for concatenated quantum Hamming codes with two to four concatenation levels under independent bit-flip errors with probability $p$. Power-law fits $p_L \propto p^{\alpha}$ are shown for each curve, where $p_L$ denotes the logical error rate and $\alpha$ indicates the error suppression exponent. The fitted exponents for the bidirectional decoder are indicated by boxed annotations; these values are close to the theoretical limits $\lceil 3^L/2 \rceil$ expected for distance-$3^L$ codes. The threshold is estimated to be $1.56\%$ for the local hard-decision decoder and $4.35\%$ for the bidirectional decoder. The bidirectional decoder not only improves the threshold by nearly $3\times$ but also achieves much steeper error suppression (higher $\alpha$) than the local decoder.}
\end{figure*}

\section{Numerical results}\label{sec:results}

We evaluate the performance of our bidirectional decoder through numerical simulations, comparing it against the standard local hard-decision decoder. We employ the code-capacity noise model, where independent bit-flip ($X$) errors occur on each physical qubit with probability $p$, and syndrome measurements are assumed to be perfect. For a given physical error rate $p$ and code instance, we estimate the logical error rate $p_{\mathrm{L}}$---defined as the probability that at least one logical qubit is incorrectly decoded---using Monte Carlo sampling. The number of trials is chosen adaptively, terminating once at least $300$ logical error events have been collected.
In the sub-threshold regime (small $p$), the logical error rate follows a power law $p_{\mathrm{L}} \propto p^{\alpha}$, where the exponent $\alpha$ characterizes the effective distance of the code. For a code with distance $d$ decoded optimally, $\alpha$ is expected to approach $\lceil d/2 \rceil$.

\subsection{Local vs. bidirectional hard-decision decoding}

Fig.~\ref{fig:hamming1a2} compares the logical error rate scaling of our bidirectional decoder with that of the local hard-decision decoder for concatenated $[[15,7,3]]$ quantum Hamming codes. The simulation extends to four concatenation levels, corresponding to a block size of $15^4 = 50,625$ physical qubits.

Our results demonstrate a significant improvement in the threshold. While the local hard-decision decoder lacks a clear threshold crossing---likely due to finite-size effects---the intersection points for three- and four-level concatenations suggest a pseudo-threshold around $p_{\textrm{th}} \approx 1.56\%$. In contrast, the bidirectional decoder exhibits a distinct threshold at approximately $p_{\textrm{th}} \approx 4.35\%$, a nearly threefold improvement.

Crucially, the bidirectional decoder also yields a substantially steeper scaling of the logical error rate. For an $L$-level CQHC under local decoding, the minimum-weight uncorrectable error has weight $2^L$, which is strictly less than the code distance $\lceil 3^L/2 \rceil$. Consistent with this, our fitted exponents $\alpha$ for the local decoder remain close to $2^L$. By comparison, the bidirectional decoder achieves exponents of $\alpha \approx 5.7, 15.3, \text{ and } 39.2$ for $L=2, 3, \text{ and } 4$, respectively. These values align closely with the theoretical limits $\lceil 3^L/2 \rceil = 5, 14, 41$, indicating effective distance preservation.

To quantify the performance gain in the low-error regime, we compare the two decoders at an equal logical capacity of $7^4=2401$ logical qubits. We consider a system of seven level-3 CQHC blocks decoded bidirectionally versus a single level-4 CQHC block decoded locally. We extrapolate the logical error rates to a physical error rate of $p=1\%$ using the ansatz $p_L \approx A (p/p_{\mathrm{th}})^\alpha$.
For the level-3 CQHC using the bidirectional decoder, the fit parameters $p_{\mathrm{th}} \approx 4.35\%$, $\alpha \approx 15.3$, and $A \approx 0.35$ yield a logical error rate of
\[
	p_L \approx 0.35 \left(\frac{p}{0.0435}\right)^{15.3} \approx 
	5.96 \times 10^{-11}.
\]
The aggregate error rate for seven blocks is thus approximately $7 \times 5.96 \times 10^{-11} \approx 4.2 \times 10^{-10}$.
In contrast, for the level-4 CQHC using the local decoder, the fit parameters $p_{\mathrm{th}} \approx 1.56\%$, $\alpha \approx 14.7$, and $A \approx 0.12$ result in
\[
	p_L \approx 0.12 \left(\frac{p}{0.0156}\right)^{14.7} \approx 1.7 \times 10^{-4}.
\]
Consequently, the bidirectional decoder outperforms the local decoder by a factor of roughly $4 \times 10^5$, while using less than half the physical qubits, specifically $23,625$ compared to $50,625$.

We note that the fitted exponents for $L=2$ and $3$ slightly exceed the theoretical bounds. This suggests the presence of an error floor at lower physical error rates, which would require rare-event sampling techniques to resolve~\cite{bravyi2013simulation,beverland2025failfasttechniquesprobe}. Conversely, the exponent for $L=4$ falls slightly below the theoretical maximum. While this could imply that the decoder does not strictly preserve the full distance, we have not identified any specific uncorrectable error patterns of weight less than 41. Practically, the observed $\alpha$ is sufficiently high that the precise distance scaling becomes secondary to the encoding rate.

\subsection{Concatenations with heterogeneous quantum Hamming codes}

\begin{figure}[t]
	\centering
	\includegraphics[width=\linewidth]{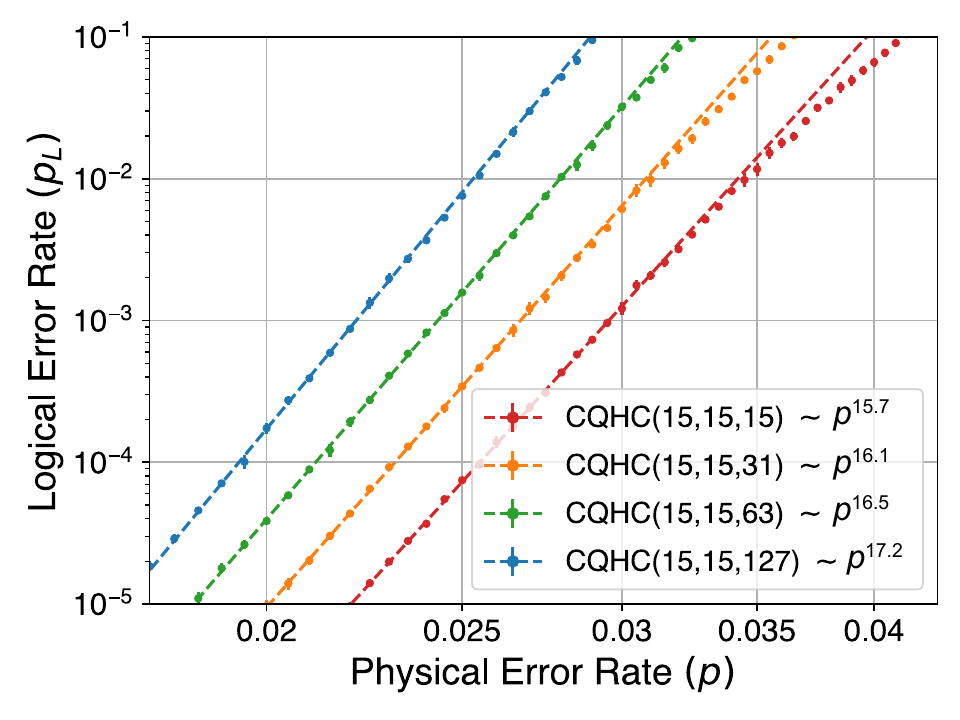}
	\caption{Logical error rates for $\mathrm{CQHC}(15,15,n)$ with $n = 15, 31, 63, 127$. In all cases, the level-1 and level-2 local codes are fixed to the $[[15,7,3]]$ quantum Hamming code, while the level-3 local quantum Hamming code determines both the encoding rate and the resulting decoding performance. Although increasing the top-level block size $n$ slightly increases the logical error rate, it significantly boosts the encoding rate while maintaining a high effective distance (slope $\alpha \approx 16$), demonstrating the flexibility of the architecture.}
\end{figure}

The bidirectional decoder naturally extends to heterogeneous concatenations. We denote a heterogeneous $L$-level concatenated code as $\mathrm{CQHC}(n_1, n_2, \dots, n_L)$, where $n_\ell$ is the block length of the local QHC at level $\ell$.
As a proof-of-principle demonstration, we benchmark the decoder on three-level concatenated quantum Hamming codes of the form $\mathrm{CQHC}(15,15,n)$ with $n \in \{15, 31, 63, 127\}$. Here, the first two concatenation levels are fixed to the $[[15,7,3]]$ code, while the highest level is a QHC of block length $n$.
We find that the fitted exponent $\alpha$ for $\mathrm{CQHC}(15,15,n)$ ranges from $15.7$ to $17.2$, consistently exceeding the theoretical value of $14$. While selecting a larger block length $n$ increases the logical error rate, it also improves the overall encoding rate. This trade-off enables more flexible design of concatenated Hamming-code architectures by balancing encoding efficiency against logical error performance.

\begin{figure}[t]
	\centering
	\includegraphics[width=\linewidth]{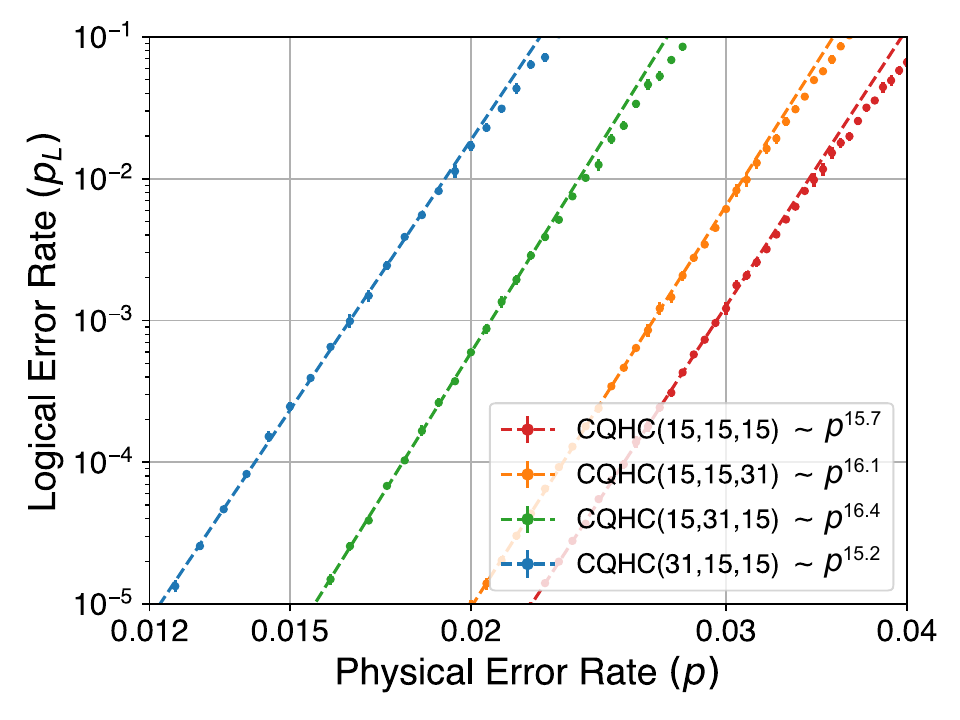}
	\caption{\label{fig:hamming4_445}Logical error rates for concatenated quantum Hamming codes with heterogeneous structures. The codes $\mathrm{CQHC}(15,15,31)$, $\mathrm{CQHC}(15,31,15)$, and $\mathrm{CQHC}(31,15,15)$ each encode 1029 logical qubits into 6975 physical qubits by concatenating two $[[15,7,3]]$ QHCs with one $[[31,21,3]]$ QHC, arranged at different concatenation levels. The configuration $\mathrm{CQHC}(15,15,31)$, which places the largest code block at the highest level, achieves the lowest logical error rate. This indicates that to maximize performance in heterogeneous concatenations, larger code blocks should be prioritized for higher concatenation levels.}
\end{figure}

It is instructive to further examine how the ordering of heterogeneous code blocks within a concatenation affects overall performance. Figure~\ref{fig:hamming4_445} compares several concatenated codes that encode the same number of logical qubits but differ in their concatenation structure. The constructions $\mathrm{CQHC}(31,15,15)$, $\mathrm{CQHC}(15,31,15)$, and $\mathrm{CQHC}(15,15,31)$ use two $[[15,7,3]]$ codes and one $[[31,21,3]]$ code, with the larger block placed at level $1$, $2$ or $3$, respectively. We find that $\mathrm{CQHC}(15,15,31)$ achieves the lowest logical error rate among these constructions, indicating that larger blocks should be placed at higher levels instead of lower levels. We expect that this conclusion will persist under circuit-level noise, since the fault-tolerant circuits for smaller code blocks are typically simpler.
\section{Discussion}\label{sec:discussion}

The bidirectional decoder developed in this work relies on several structural
properties of CQHCs that allow the
reassignment of logical flips to be implemented greedily with polynomial-time
complexity.
In this section, we clarify these assumptions, which delineate the conditions
under which the bidirectional decoder is valid and indicate how it must be
modified when applied to other concatenated code families.

A first assumption concerns the \textsf{FlipCost} subroutine.
This subroutine assumes that a level-$\ell$ logical flip $\Delta$ admits a
\emph{unique} representation at level $(\ell-1)$ that is supported on the minimum
number of level-$(\ell-1)$ subblocks.
This typically requires that each minimum-weight logical operator of the local
code at level $\ell$ has a unique minimum-weight physical representation.
This assumption is violated, for example, when the local code is chosen to be the
\([[7,1,3]]\) Steane code, for which the logical $X$ operator has multiple distinct
weight-$3$ physical implementations.
In this case, \textsf{FlipCost} selects only a single representation and does not
treat all minimum-weight realizations on equal footing.
When propagated to lower concatenation levels, these different representations
can lead to different physical costs, since lower-level recoveries differ and
these differences are not visible at level $\ell-1$.
As a result, the greedy reassignment may fail to identify the true minimum-cost
implementation of the logical flip, effectively reducing the achievable code
distance.
An explicit example of this failure mode for a three-level concatenated
\([[7,1,3]]\) Steane code is provided in Appendix~\ref{appendix:concat713}.

For larger quantum Hamming codes, this ambiguity does not arise.
All weight-$3$ logical operators have unique weight-$3$ representations, since
all stabilizer elements have weight strictly larger than $6$, and multiplying a
logical operator by any stabilizer necessarily increases its weight.
Moreover, concatenating even a single level of the Steane code reduces the
encoding rate by a factor of $7$, making such constructions less favorable in the
high-rate regime considered here.
If one nevertheless wishes to use the Steane code at lower concatenation levels,
for instance to improve the threshold, the \textsf{FlipCost} subroutine and the
decoder at those levels can be replaced by Poulin’s maximum-likelihood decoder~\cite{Poulin2006}.

A second relevant property is that CQHCs have odd distance.
At each concatenation level, this largely excludes the possibility that an error
admits multiple minimum-weight recovery candidates.
By contrast, for concatenations of even-distance codes, such as many-hypercube
codes with distance $2^L$, errors of weight $d/2$ necessarily admit multiple
minimum-weight recoveries.
If only a single recovery candidate is retained for each level-$(\ell-1)$
subblock, the bidirectional decoder may fail to preserve distance.
In particular, there may exist combinations of subblock recoveries for which the
induced level-$\ell$ syndrome is trivial, while other combinations yield a
nontrivial level-$\ell$ syndrome.
Selecting an incompatible subblock recovery can therefore prevent the decoder
from identifying the minimum-weight recovery for the level-$\ell$ code.
In general, all combinations of minimum-weight subblock recoveries must be
considered, leading to a substantial increase in algorithmic complexity.
This situation is analogous to the ``split belief'' phenomenon~\cite{poulin2008iterative} encountered in
belief-propagation decoding of quantum LDPC codes, where multiple locally
consistent decisions coexist and additional post-processing, such as
order-statistics decoding~\cite{Panteleev2021degeneratequantum,Roffe2020}, is required to resolve the ambiguity.

Relatedly, when multiple minimum-weight recovery candidates exist, the
reassignment subroutine cannot be executed in a fully greedy manner.
For example, if transferring a logical flip between subblocks yields equal
physical cost, all such transfers may correspond to valid minimum-weight
reassignments and should be retained.
Accounting for these possibilities can result in an exponentially large search
space of candidate reassignments.
For certain code families, such as many-hypercube codes, this search space remains
small, since the logical flip can be assumed to act on a single block, reducing
the enumeration to the size of the local code block~\cite{Goto2024}.
For larger codes, however, such enumeration may become prohibitive.

In fact, even for CQHCs, errors with multiple minimum-weight recovery candidates
can occur.
As an example, in a two-level CQHC there exists a weight-$12$ logical operator
\[
	\{(a,b)_0 : a \in \{2,3,4,5\},\ b \in \{1,2,3\}\},
\]
which can be decomposed into two disjoint error patterns of equal weight,
\[
	E_1 = \{(2,1)_0,(3,1)_0,(4,1)_0,(5,1)_0,(2,2)_0,(3,2)_0\}
\]
and
\[
	E_2 = \{(4,2)_0,(5,2)_0,(2,3)_0,(3,3)_0,(4,3)_0,(5,3)_0\}.
\]
Neither $E_1$ nor $E_2$ admits a recovery of weight smaller than $6$.
Nevertheless, the probability of such error patterns is significantly suppressed,
and their impact on the typical performance of the decoder is limited.

\section{Conclusions and outlook}\label{sec:conclusion}

In this work, we introduced a hard-decision decoder that substantially improves the
decoding performance of high-rate concatenated quantum Hamming codes.
The bidirectional flow of information enables the decoder to achieve polynomial-time
worst-case computational complexity while overcoming limitations of traditional
local hard-decision decoding.
By mitigating correctable error patterns that are otherwise miscorrected, the
decoder significantly enhances the effective code distance, bringing performance
closer to the theoretical limits.

An immediate direction for future work is the extension of this approach to more
realistic circuit-level noise models.
For concatenated quantum codes, Steane- or Knill-style syndrome extraction~\cite{Steane1997active,knill2005quantum} is
expected to be advantageous, as these methods minimize back-action on the data.
Assuming that auxiliary logical resource states, such as logical Bell states, can be
prepared fault-tolerantly, one can apply a code-capacity-level decoder directly to
the extracted syndrome without explicitly accounting for errors on individual
syndrome bits.
Realizing this assumption at the circuit level, however, requires substantial
optimization of fault-tolerant state-preparation, verification, and postselection
protocols.
The design and optimization of such circuits constitute a significant body of work
and are beyond the scope of the present study.

Another promising direction is the exploration of hybrid concatenation schemes.
For example, one may concatenate quantum Hamming codes with distance-$2$ Iceberg
codes~\cite{rains2002quantum} to obtain architectures that interpolate between CQHCs and
many-hypercube codes.
Although Iceberg codes are error-detection codes, they provide greater flexibility
in encoding rate and block size than quantum Hamming codes.
More broadly, it would be interesting to study concatenations involving other small
distance-$3$ stabilizer codes, both CSS and non-CSS, and to determine which of these
codes admit efficient decoding within the bidirectional framework developed here.
Such investigations will substantially expand the space of local codes suitable
for concatenated constructions.

Finally, we note that the proposed bidirectional decoder is amenable to parallelization. The recursive structure of the algorithm allows independent subblocks to be decoded simultaneously. Additionally, certain components of the reassignment subroutine, such as the evaluation of physical costs for different subblocks, can be distributed across multiple processing units. Although the greedy heuristic inherently involves sequential dependencies, specific steps such as the exploration of different initial stabilizer choices can be parallelized. Developing a variant of the greedy strategy that admits full parallelization would be an interesting direction for future research. 

\newline
\section*{Acknowledgement}
We thank Sirui Lu for introducing TeXRA and for technical support during manuscript preparation.
This work is supported under the Guangdong Provincial Quantum Science Strategic Initiative (Grants No. GDZX2503001).

\bibliography{Reference}

@article{pino_demonstration_2021,
	title = {Demonstration of the trapped-ion quantum {CCD} computer architecture},
	volume = {592},
	copyright = {2021 The Author(s), under exclusive licence to Springer Nature Limited part of Springer Nature},
	issn = {1476-4687},
	url = {https://www.nature.com/articles/s41586-021-03318-4},
	doi = {10.1038/s41586-021-03318-4},
	language = {en},
	number = {7853},
	urldate = {2026-01-12},
	journal = {Nature},
	author = {Pino, J. M. and Dreiling, J. M. and Figgatt, C. and Gaebler, J. P. and Moses, S. A. and Allman, M. S. and Baldwin, C. H. and Foss-Feig, M. and Hayes, D. and Mayer, K. and Ryan-Anderson, C. and Neyenhuis, B.},
	month = apr,
	year = {2021},
	pages = {209--213},
}

@article{main_distributed_2025,
	title = {Distributed quantum computing across an optical network link},
	volume = {638},
	issn = {1476-4687},
	url = {https://doi.org/10.1038/s41586-024-08404-x},
	doi = {10.1038/s41586-024-08404-x},
	number = {8050},
	journal = {Nature},
	author = {Main, D. and Drmota, P. and Nadlinger, D. P. and Ainley, E. M. and Agrawal, A. and Nichol, B. C. and Srinivas, R. and Araneda, G. and Lucas, D. M.},
	month = feb,
	year = {2025},
	pages = {383--388},
}

@article{almanakly_deterministic_2025,
	title = {Deterministic remote entanglement using a chiral quantum interconnect},
	volume = {21},
	issn = {1745-2481},
	url = {https://doi.org/10.1038/s41567-025-02811-1},
	doi = {10.1038/s41567-025-02811-1},
	number = {5},
	journal = {Nature Physics},
	author = {Almanakly, Aziza and Yankelevich, Beatriz and Hays, Max and Kannan, Bharath and Assouly, Réouven and Greene, Alex and Gingras, Michael and Niedzielski, Bethany M. and Stickler, Hannah and Schwartz, Mollie E. and Serniak, Kyle and Wang, Joel \^{I}-j. and Orlando, Terry P. and Gustavsson, Simon and Grover, Jeffrey A. and Oliver, William D.},
	month = may,
	year = {2025},
	pages = {825--830},
}

@article{sahu_entangling_2023,
	title = {Entangling microwaves with light},
	volume = {380},
	url = {https://doi.org/10.1126/science.adg3812},
	doi = {10.1126/science.adg3812},
	number = {6646},
	urldate = {2026-01-12},
	journal = {Science},
	author = {Sahu, R. and Qiu, L. and Hease, W. and Arnold, G. and Minoguchi, Y. and Rabl, P. and Fink, J. M.},
	month = may,
	year = {2023},
	pages = {718--721},
}

@INPROCEEDINGS{shor1996fault,
  author={Shor, P.W.},
  booktitle={Proceedings of 37th Conference on Foundations of Computer Science}, 
  title={Fault-tolerant quantum computation}, 
  year={1996},
  volume={},
  number={},
  pages={56-65},
  url = {http://dx.doi.org/10.1109/sfcs.1996.548464},
  publisher = {IEEE Comput. Soc. Press},
  doi={10.1109/SFCS.1996.548464}
}

@article{gottesman1998theory,
  title = {Theory of fault-tolerant quantum computation},
  author = {Gottesman, Daniel},
  journal = {Phys. Rev. A},
  volume = {57},
  issue = {1},
  pages = {127--137},
  numpages = {0},
  year = {1998},
  month = {Jan},
  publisher = {American Physical Society},
  doi = {10.1103/PhysRevA.57.127},
  url = {https://link.aps.org/doi/10.1103/PhysRevA.57.127}
}

@misc{bravyi1998quantum,
      title={Quantum codes on a lattice with boundary}, 
      author={S. B. Bravyi and A. Yu. Kitaev},
      year={1998},
      howpublished = {\href{https://arxiv.org/abs/quant-ph/9811052}{arXiv:quant-ph/981105}},
}

@article{dennis2002topological,
    author = {Dennis, Eric and Kitaev, Alexei and Landahl, Andrew and Preskill, John},
    title = {Topological quantum memory},
    publisher = {AIP Publishing},
    journal = {J. Math. Phys.},
    volume = {43},
    number = {9},
    pages = {4452-4505},
    year = {2002},
    month = {09},
    issn = {0022-2488},
    doi = {10.1063/1.1499754},
    url = {https://doi.org/10.1063/1.1499754}
}

@article{fowler2009high,
  title = {High-threshold universal quantum computation on the surface code},
  author = {Fowler, Austin G. and Stephens, Ashley M. and Groszkowski, Peter},
  journal = {Phys. Rev. A},
  volume = {80},
  issue = {5},
  pages = {052312},
  numpages = {14},
  year = {2009},
  month = {Nov},
  publisher = {American Physical Society},
  doi = {10.1103/PhysRevA.80.052312},
  url = {https://link.aps.org/doi/10.1103/PhysRevA.80.052312}
}

@article{Pattison2025hierarchical,
  doi = {10.22331/q-2025-05-05-1728},
  url = {https://doi.org/10.22331/q-2025-05-05-1728},
  title = {Hierarchical memories: {S}imulating quantum {LDPC} codes with local gates},
  author = {Pattison, Christopher A. and Krishna, Anirudh and Preskill, John},
  journal = {{Quantum}},
  issn = {2521-327X},
  publisher = {{Verein zur F{\"{o}}rderung des Open Access Publizierens in den Quantenwissenschaften}},
  volume = {9},
  pages = {1728},
  month = may,
  year = {2025}
}

@article{Panteleev2021degeneratequantum,
  doi = {10.22331/q-2021-11-22-585},
  url = {https://doi.org/10.22331/q-2021-11-22-585},
  title = {Degenerate {Q}uantum {LDPC} {C}odes {W}ith {G}ood {F}inite {L}ength {P}erformance},
  author = {Panteleev, Pavel and Kalachev, Gleb},
  journal = {{Quantum}},
  issn = {2521-327X},
  publisher = {{Verein zur F{\"{o}}rderung des Open Access Publizierens in den Quantenwissenschaften}},
  volume = {5},
  pages = {585},
  month = nov,
  year = {2021}
}

@article{Roffe2020,
  title = {Decoding across the quantum low-density parity-check code landscape},
  author = {Roffe, Joschka and White, David R. and Burton, Simon and Campbell, Earl},
  journal = {Phys. Rev. Res.},
  volume = {2},
  issue = {4},
  pages = {043423},
  numpages = {13},
  year = {2020},
  month = {Dec},
  publisher = {American Physical Society},
  doi = {10.1103/PhysRevResearch.2.043423},
  url = {https://link.aps.org/doi/10.1103/PhysRevResearch.2.043423}
}

@article{junichi2025,
    author = {Haruna, Junichi and Fujii, Keisuke},
    title = {Hierarchical Quantum Error Correction with Hypergraph Product Code and Rotated Surface Code},
    journal = {Prog. Theor. Exp. Phys.},
    volume = {2025},
    number = {10},
    pages = {103A03},
    year = {2025},
    month = {09},
    issn = {2050-3911},
    doi = {10.1093/ptep/ptaf130}
}

@article{gidney2025yoked,
	author = {Gidney, Craig and Newman, Michael and Brooks, Peter and Jones, Cody},
	date = {2025/05/14},
	doi = {10.1038/s41467-025-59714-1},
	id = {Gidney2025},
	isbn = {2041-1723},
	journal = {Nat. Commun.},
	number = {1},
	pages = {4498},
	title = {Yoked surface codes},
	url = {https://doi.org/10.1038/s41467-025-59714-1},
	volume = {16},
	year = {2025}}

@article{fowler2012surface,
  title = {Surface codes: Towards practical large-scale quantum computation},
  author = {Fowler, Austin G. and Mariantoni, Matteo and Martinis, John M. and Cleland, Andrew N.},
  journal = {Phys. Rev. A},
  volume = {86},
  issue = {3},
  pages = {032324},
  numpages = {48},
  year = {2012},
  month = {Sep},
  publisher = {American Physical Society},
  doi = {10.1103/PhysRevA.86.032324},
  url = {https://link.aps.org/doi/10.1103/PhysRevA.86.032324}
}

@article{google2023suppressing,
  title = {Suppressing Quantum Errors by Scaling a Surface Code Logical Qubit},
  author = {
  Acharya, Rajeev and
  Aleiner, Igor and
  Allen, Richard and
  Andersen, Trond I. and
  Ansmann, Markus and
  Arute, Frank and
  Arya, Kunal and
  Asfaw, Abraham and
  Atalaya, Juan and
  Babbush, Ryan and
  Bacon, Dave and
  Bardin, Joseph C. and
  Basso, Joao and
  Bengtsson, Andreas and
  Boixo, Sergio
  and others
  },
  collaboration = {Google Quantum AI},
  year = 2023,
  month = feb,
  journal = {Nature},
  volume = {614},
  number = {7949},
  pages = {676--681},
  issn = {1476-4687},
  doi = {10.1038/s41586-022-05434-1},
  url = {https://doi.org/10.1038/s41586-022-05434-1}
}

@article{breuckmann2021quantum,
  title = {Quantum Low-Density Parity-Check Codes},
  author = {Breuckmann, Nikolas P. and Eberhardt, Jens Niklas},
  journal = {PRX Quantum},
  volume = {2},
  issue = {4},
  pages = {040101},
  numpages = {19},
  year = {2021},
  month = {Oct},
  publisher = {American Physical Society},
  doi = {10.1103/PRXQuantum.2.040101},
  url = {https://link.aps.org/doi/10.1103/PRXQuantum.2.040101}
}

@article{knill2005quantum,
author={Knill, E.},
title={Quantum computing with realistically noisy devices},
journal={Nature},
year={2005},
month={Mar},
day={01},
volume={434},
number={7029},
pages={39-44},
issn={1476-4687},
doi={10.1038/nature03350},
url={https://doi.org/10.1038/nature03350}
}

@misc{tansuwannont2025clifford,
      title={Clifford gates with logical transversality for self-dual CSS codes}, 
      author={Theerapat Tansuwannont and Yugo Takada and Keisuke Fujii},
      year={2025},
      howpublished = {\href{https://arxiv.org/abs/2503.19790}{arXiv:2503.19790}},
}

@article{yoshida2025concatenate,
author={Yoshida, Satoshi
and Tamiya, Shiro
and Yamasaki, Hayata},
title={Concatenate codes, save qubits},
journal={npj Quantum Inf.},
year={2025},
month={May},
day={31},
volume={11},
number={1},
pages={88},
issn={2056-6387},
doi={10.1038/s41534-025-01035-8},
url={https://doi.org/10.1038/s41534-025-01035-8}
}

@article{Yamasaki2024,
author={Yamasaki, Hayata
and Koashi, Masato},
title={Time-Efficient Constant-Space-Overhead Fault-Tolerant Quantum Computation},
journal={Nat. Phys.},
year={2024},
month={Feb},
day={01},
volume={20},
number={2},
pages={247-253},
issn={1745-2481},
doi={10.1038/s41567-023-02325-8},
url={https://doi.org/10.1038/s41567-023-02325-8}
}

@article{Steane1996,
  title = {Simple quantum error-correcting codes},
  author = {Steane, A. M.},
  journal = {Phys. Rev. A},
  volume = {54},
  issue = {6},
  pages = {4741--4751},
  numpages = {0},
  year = {1996},
  month = {Dec},
  publisher = {American Physical Society},
  doi = {10.1103/PhysRevA.54.4741},
  url = {https://link.aps.org/doi/10.1103/PhysRevA.54.4741}
}

@article{Poulin2006,
  title = {Optimal and efficient decoding of concatenated quantum block codes},
  author = {Poulin, David},
  journal = {Phys. Rev. A},
  volume = {74},
  issue = {5},
  pages = {052333},
  numpages = {5},
  year = {2006},
  month = {Nov},
  publisher = {American Physical Society},
  doi = {10.1103/PhysRevA.74.052333},
  url = {https://link.aps.org/doi/10.1103/PhysRevA.74.052333}
}

@article{Goto2024,
author = {Hayato Goto },
title = {High-performance fault-tolerant quantum computing with many-hypercube codes},
journal = {Sci. Adv.},
volume = {10},
number = {36},
pages = {eadp6388},
year = {2024},
issn      = {2375-2548},
doi = {10.1126/sciadv.adp6388},
URL = {https://www.science.org/doi/abs/10.1126/sciadv.adp6388}
}

@article{wilde2009logical,
  title = {Logical operators of quantum codes},
  author = {Wilde, Mark M.},
  journal = {Phys. Rev. A},
  volume = {79},
  issue = {6},
  pages = {062322},
  numpages = {5},
  year = {2009},
  month = {Jun},
  publisher = {American Physical Society},
  doi = {10.1103/PhysRevA.79.062322},
  url = {https://link.aps.org/doi/10.1103/PhysRevA.79.062322}
}

@techreport{knill1996concatenated,
  title       = {Concatenated Quantum Codes},
  author      = {Knill, Emanuel and Laflamme, Raymond},
  institution = {Office of Scientific and Technical Information},
  year        = {1996},
  month       = {jul},
  doi         = {10.2172/369608},
  url         = {https://doi.org/10.2172/369608}
}

@article{tillich2013quantum,
  author={Tillich, Jean-Pierre and Zémor, Gilles},
  journal={IEEE Trans. Inf. Theory},
  title={Quantum LDPC Codes With Positive Rate and Minimum Distance Proportional to the Square Root of the Blocklength}, 
  year={2014},
  month     = {feb},
  volume={60},
  number={2},
  pages={1193-1202},
  issn      = {1557-9654},
  doi={10.1109/TIT.2013.2292061},
  url= {http://dx.doi.org/10.1109/tit.2013.2292061},
}

@ARTICLE{panteleev2021quantum,
  author={Panteleev, Pavel and Kalachev, Gleb},
  journal={IEEE Trans. Inf. Theory},
  title={Quantum LDPC Codes With Almost Linear Minimum Distance}, 
  year={2022},
  month     = {jan},
  volume={68},
  number={1},
  pages={213-229},
  doi={10.1109/TIT.2021.3119384},
  url       = {http://dx.doi.org/10.1109/tit.2021.3119384}
}

@article{shor1995scheme,
  title = {Scheme for reducing decoherence in quantum computer memory},
  author = {Shor, Peter W.},
  journal = {Phys. Rev. A},
  volume = {52},
  issue = {4},
  pages = {R2493--R2496},
  numpages = {0},
  year = {1995},
  month = {Oct},
  publisher = {American Physical Society},
  doi = {10.1103/PhysRevA.52.R2493},
  url = {https://link.aps.org/doi/10.1103/PhysRevA.52.R2493}
}

@article{steane1996error,
  title = {Error Correcting Codes in Quantum Theory},
  author = {Steane, A. M.},
  journal = {Phys. Rev. Lett.},
  volume = {77},
  issue = {5},
  pages = {793--797},
  numpages = {0},
  year = {1996},
  month = {Jul},
  publisher = {American Physical Society},
  doi = {10.1103/PhysRevLett.77.793},
  url = {https://link.aps.org/doi/10.1103/PhysRevLett.77.793}
}

@misc{gottesman1997stabilizer,
  author = {Gottesman, Daniel},
  title  = {Stabilizer Codes and Quantum Error Correction},
  year   = {1997},
  howpublished = {\href{https://arxiv.org/abs/quant-ph/9705052}{arXiv:quant-ph/9705052}}
}

@article{knill1997theory,
  title = {Theory of quantum error-correcting codes},
  author = {Knill, Emanuel and Laflamme, Raymond},
  journal = {Phys. Rev. A},
  volume = {55},
  issue = {2},
  pages = {900--911},
  numpages = {0},
  year = {1997},
  month = {Feb},
  publisher = {American Physical Society},
  doi = {10.1103/PhysRevA.55.900},
  url = {https://link.aps.org/doi/10.1103/PhysRevA.55.900}
}

@inproceedings{aharonov1997fault,
author = {Aharonov, D. and Ben-Or, M.},
title = {Fault-tolerant quantum computation with constant error},
year = {1997},
isbn = {0897918886},
publisher = {Association for Computing Machinery},
address = {New York, NY, USA},
url = {https://doi.org/10.1145/258533.258579},
doi = {10.1145/258533.258579},
booktitle = {Proceedings of the Twenty-Ninth Annual ACM Symposium on Theory of Computing},
pages = {176–188},
numpages = {13},
location = {El Paso, Texas, USA},
series = {STOC '97}
}

@article{preskill1998reliable,
    author = {Preskill, John},
    title = {Reliable quantum computers},
    journal={Proc. R. Soc. London, Ser. A},
    volume = {454},
    number = {1969},
    pages = {385-410},
    year = {1998},
    month = {01},
    issn = {1364-5021},
    doi = {10.1098/rspa.1998.0167},
    url = {https://doi.org/10.1098/rspa.1998.0167}
}

@article{stephens2014fault,
  title = {Fault-tolerant thresholds for quantum error correction with the surface code},
  author = {Stephens, Ashley M.},
  journal = {Phys. Rev. A},
  volume = {89},
  issue = {2},
  pages = {022321},
  numpages = {9},
  year = {2014},
  month = {Feb},
  publisher = {American Physical Society},
  doi = {10.1103/PhysRevA.89.022321},
  url = {https://link.aps.org/doi/10.1103/PhysRevA.89.022321}
}

@article{kitaev2003fault,
title = {Fault-tolerant quantum computation by anyons},
journal = {Annals of Physics},
volume = {303},
number = {1},
pages = {2-30},
year = {2003},
issn = {0003-4916},
doi = {https://doi.org/10.1016/S0003-4916(02)00018-0},
url = {https://www.sciencedirect.com/science/article/pii/S0003491602000180},
author = {A.Yu. Kitaev}
}

@article{barends2014superconducting,
  title     = {Superconducting quantum circuits at the surface code threshold for fault tolerance},
  volume    = {508},
  issn      = {1476-4687},
  url       = {http://dx.doi.org/10.1038/nature13171},
  doi       = {10.1038/nature13171},
  number    = {7497},
  journal   = {Nature},
  publisher = {Springer Science and Business Media LLC},
  author    = {Barends, R. and Kelly, J. and Megrant, A. and Veitia, A. and Sank, D. and Jeffrey, E. and White, T. C. and Mutus, J. and Fowler, A. G. and Campbell, B. and Chen, Y. and Chen, Z. and Chiaro, B. and Dunsworth, A. and Neill, C. and O’Malley, P. and Roushan, P. and Vainsencher, A. and Wenner, J. and Korotkov, A. N. and Cleland, A. N. and Martinis, John M.},
  year      = {2014},
  month={Apr},
  day={01},
  pages     = {500–503}
}

@article{zhao2022realization,
  title = {Realization of an Error-Correcting Surface Code with Superconducting Qubits},
  author = {Zhao, Youwei and Ye, Yangsen and Huang, He-Liang and Zhang, Yiming and Wu, Dachao and Guan, Huijie and Zhu, Qingling and Wei, Zuolin and He, Tan and Cao, Sirui and Chen, Fusheng and Chung, Tung-Hsun and Deng, Hui and Fan, Daojin and Gong, Ming and Guo, Cheng and Guo, Shaojun and Han, Lianchen and Li, Na and Li, Shaowei and Li, Yuan and Liang, Futian and Lin, Jin and Qian, Haoran and Rong, Hao and Su, Hong and Sun, Lihua and Wang, Shiyu and Wu, Yulin and Xu, Yu and Ying, Chong and Yu, Jiale and Zha, Chen and Zhang, Kaili and Huo, Yong-Heng and Lu, Chao-Yang and Peng, Cheng-Zhi and Zhu, Xiaobo and Pan, Jian-Wei},
  journal = {Phys. Rev. Lett.},
  volume = {129},
  issue = {3},
  pages = {030501},
  numpages = {7},
  year = {2022},
  month = {Jul},
  publisher = {American Physical Society},
  doi = {10.1103/PhysRevLett.129.030501},
  url = {https://link.aps.org/doi/10.1103/PhysRevLett.129.030501}
}

@article{bluvstein2024logical,
author={Bluvstein, Dolev
and Evered, Simon J.
and Geim, Alexandra A.
and Li, Sophie H.
and Zhou, Hengyun
and Manovitz, Tom
and Ebadi, Sepehr
and Cain, Madelyn
and Kalinowski, Marcin
and Hangleiter, Dominik
and Bonilla Ataides, J. Pablo
and Maskara, Nishad
and Cong, Iris
and Gao, Xun
and Sales Rodriguez, Pedro
and Karolyshyn, Thomas
and Semeghini, Giulia
and Gullans, Michael J.
and Greiner, Markus
and Vuleti{\'{c}}, Vladan
and Lukin, Mikhail D.},
title={Logical quantum processor based on reconfigurable atom arrays},
journal={Nature},
year={2024},
month={Feb},
day={01},
volume={626},
number={7997},
pages={58-65},
issn={1476-4687},
doi={10.1038/s41586-023-06927-3},
url={https://doi.org/10.1038/s41586-023-06927-3}
}

@article{knill1998resilient,
    author = {Knill, Emanuel and Laflamme, Raymond and Zurek, Wojciech H.},
    title = {Resilient quantum computation: error models and thresholds},
    journal={Proc. R. Soc. London, Ser. A},
    volume = {454},
    number = {1969},
    pages = {365-384},
    year = {1998},
    month = {01},
    issn = {1364-5021},
    doi = {10.1098/rspa.1998.0166},
    url = {https://doi.org/10.1098/rspa.1998.0166}
}

@article{aliferis2005quantum,
author = {Aliferis, Panos and Gottesman, Daniel and Preskill, John},
title = {Quantum accuracy threshold for concatenated distance-3 codes},
year = {2006},
issue_date = {March 2006},
publisher = {Rinton Press, Incorporated},
address = {Paramus, NJ},
volume = {6},
number = {2},
issn = {1533-7146},
journal = {Quantum Inf. Comput.},
month = mar,
pages = {97–165},
numpages = {69}
}

@article{rains2002quantum,
  title     = {Quantum codes of minimum distance two},
  volume    = {45},
  issn      = {0018-9448},
  url       = {http://dx.doi.org/10.1109/18.746807},
  doi       = {10.1109/18.746807},
  number    = {1},
  journal = {IEEE Trans. Inf. Theory},
  publisher = {Institute of Electrical and Electronics Engineers (IEEE)},
  author    = {Rains, E.M.},
  year      = {1999},
  pages     = {266–271}
}

@article{Baspin2022,
  doi = {10.22331/q-2022-05-13-711},
  url = {https://doi.org/10.22331/q-2022-05-13-711},
  title = {Connectivity constrains quantum codes},
  author = {Baspin, Nou{\'{e}}dyn and Krishna, Anirudh},
  journal = {{Quantum}},
  issn = {2521-327X},
  publisher = {{Verein zur F{\"{o}}rderung des Open Access Publizierens in den Quantenwissenschaften}},
  volume = {6},
  pages = {711},
  month = may,
  year = {2022}
}

@article{Bravyi2009,
doi = {10.1088/1367-2630/11/4/043029},
url = {https://doi.org/10.1088/1367-2630/11/4/043029},
year = {2009},
month = {apr},
publisher = {},
volume = {11},
number = {4},
pages = {043029},
author = {Bravyi, Sergey and Terhal, Barbara},
title = {A no-go theorem for a two-dimensional self-correcting quantum memory based on stabilizer codes},
journal = {	New J. Phys.}
}

@article{Bravyi2010,
  title = {Tradeoffs for Reliable Quantum Information Storage in 2D Systems},
  author = {Bravyi, Sergey and Poulin, David and Terhal, Barbara},
  journal = {Phys. Rev. Lett.},
  volume = {104},
  issue = {5},
  pages = {050503},
  numpages = {4},
  year = {2010},
  month = {Feb},
  publisher = {American Physical Society},
  doi = {10.1103/PhysRevLett.104.050503},
  url = {https://link.aps.org/doi/10.1103/PhysRevLett.104.050503}
}

@article{Calderbank_1997,
  title = {Quantum Error Correction and Orthogonal Geometry},
  author = {Calderbank, A. R. and Rains, E. M. and Shor, P. W. and Sloane, N. J. A.},
  journal = {Phys. Rev. Lett.},
  volume = {78},
  issue = {3},
  pages = {405--408},
  numpages = {0},
  year = {1997},
  month = {Jan},
  publisher = {American Physical Society},
  doi = {10.1103/PhysRevLett.78.405},
  url = {https://link.aps.org/doi/10.1103/PhysRevLett.78.405}
}

@article{Calderbank1996GoodQuantum,
  title = {Good quantum error-correcting codes exist},
  author = {Calderbank, A. R. and Shor, Peter W.},
  journal = {Phys. Rev. A},
  volume = {54},
  issue = {2},
  pages = {1098--1105},
  numpages = {0},
  year = {1996},
  month = {Aug},
  publisher = {American Physical Society},
  doi = {10.1103/PhysRevA.54.1098},
  url = {https://link.aps.org/doi/10.1103/PhysRevA.54.1098}
}

@article{Steane1996Multiple,
    author = {Steane, Andrew},
    title = {Multiple-particle interference and quantum error correction},
    journal={Proc. R. Soc. London, Ser. A},
    volume = {452},
    number = {1954},
    pages = {2551-2577},
    year = {1996},
    month = {11},
    issn = {1364-5021},
    doi = {10.1098/rspa.1996.0136},
    url = {https://doi.org/10.1098/rspa.1996.0136}
}

@article{bravyi2013simulation,
  title = {Simulation of rare events in quantum error correction},
  author = {Bravyi, Sergey and Vargo, Alexander},
  journal = {Phys. Rev. A},
  volume = {88},
  issue = {6},
  pages = {062308},
  numpages = {14},
  year = {2013},
  month = {Dec},
  publisher = {American Physical Society},
  doi = {10.1103/PhysRevA.88.062308},
  url = {https://link.aps.org/doi/10.1103/PhysRevA.88.062308}
}

@misc{beverland2025failfasttechniquesprobe,
      title={Fail fast: techniques to probe rare events in quantum error correction}, 
      author={Michael E. Beverland and Malcolm Carroll and Andrew W. Cross and Theodore J. Yoder},
      year={2025},
      howpublished={\href{https://arxiv.org/abs/2511.15177}{arXiv:2511.15177}},
}

@article{poulin2008iterative,
author = {Poulin, David and Chung, Yeojin},
title = {On the iterative decoding of sparse quantum codes},
year = {2008},
issue_date = {November 2008},
publisher = {Rinton Press, Incorporated},
address = {Paramus, NJ},
volume = {8},
number = {10},
issn = {1533-7146},
journal = {Quantum Inf. Comput.},
month = {nov},
pages = {987–1000},
numpages = {14},
url={https://dl.acm.org/doi/10.5555/2016985.2016993},
}

@article{Steane1997active,
  title = {Active Stabilization, Quantum Computation, and Quantum State Synthesis},
  author = {Steane, A. M.},
  journal = {Phys. Rev. Lett.},
  volume = {78},
  issue = {11},
  pages = {2252--2255},
  numpages = {0},
  year = {1997},
  month = {Mar},
  publisher = {American Physical Society},
  doi = {10.1103/PhysRevLett.78.2252},
  url = {https://link.aps.org/doi/10.1103/PhysRevLett.78.2252}
}

\twocolumngrid
\onecolumngrid

\appendix
\section{Failure of the bidirectional decoder for a 3-level concatenated \texorpdfstring{$[[7,1,3]]$}{[[7,1,3]]} code}
\label{appendix:concat713}

In this section, we present an explicit weight-$10$ error pattern on a
3-level concatenated $[[7,1,3]]$ Steane code that is not corrected by the
bidirectional decoder.
The errors are concentrated on two level-$2$ subblocks,
denoted $\mathcal{C}^{(2)}_{1}$ and $\mathcal{C}^{(2)}_{2}$, each containing five
physical errors arranged in the same pattern.

For each $i \in \{1,2\}$, the physical errors on $\mathcal{C}^{(2)}_{i}$ occur on
the level-$0$ qubits
\begin{equation}
	\label{eq:error-pattern}
	\bigl\{(i,2,1)_0,\ (i,4,2)_0,\ (i,4,3)_0,\ (i,6,2)_0,\ (i,6,3)_0 \bigr\}.
\end{equation}
That is, the level-$1$ subblock $\mathcal{C}^{(1)}_{i,2}$ contains a single error
on its first physical qubit $(i,2,1)_0$, while for $j \in \{4,6\}$ the level-$1$ subblocks
$\mathcal{C}^{(1)}_{i,j}$ each contain two physical errors on $(i,j,2)_0$ and $(i,j,3)_0$.

Since the Steane code encodes a single logical qubit, we simplify notation by
labeling level-$1$ qubits as $(i,j)_1$ instead of $(i,j,1)_1$, and level-$2$
qubits as $(i)_2$ instead of $(i,1,1)_2$.

Before analyzing the decoding process in detail, we summarize the origin of the
failure.
After local decoding at levels~$1$ and~$2$, the decoder applies a weight-$4$ recovery to each of the subblocks $\mathcal{C}^{(2)}_{1}$ and $\mathcal{C}^{(2)}_{2}$, thereby introducing logical errors on both subblocks and producing a nontrivial level-$3$ syndrome.
The baseline recovery proposed by the bidirectional decoder is to apply a
logical $X$ flip on the third level-$2$ subblock $\mathcal{C}^{(2)}_{3}$, with a
total physical cost $9+4+4=17$.
The decoder then considers transferring this logical flip from
$\mathcal{C}^{(2)}_{3}$ to $\mathcal{C}^{(2)}_{1}$ and $\mathcal{C}^{(2)}_{2}$ via
the \textsf{Reassign} subroutine.
Although such a transfer ideally admits a physical realization of total weight $10$,
the subroutine \textsf{FlipCost} returns an estimate of weight $18$ because it
selects an unfavorable stabilizer representative that has the same
intermediate-level weight as the favorable one but a larger physical cost.
As a result, the transfer is rejected and a logical error occurs.
We now explain this process step by step.

\subsection*{Level-1 decoding}

We first consider the action of the local hard-decision decoder at level~$1$.
For $i \in \{1,2\}$ and $j \in \{2,4,6\}$, the decoder applies a recovery to each
subblock $\mathcal{C}^{(1)}_{i,j}$ on its first physical qubit, namely the
level-$0$ qubit $(i,j,1)_0$.

As a consequence, the subblocks $\mathcal{C}^{(1)}_{i,4}$ and
$\mathcal{C}^{(1)}_{i,6}$ each incur a logical error, whereas
$\mathcal{C}^{(1)}_{i,2}$ is successfully corrected.
Thus, after level-$1$ decoding, each of the two level-$2$ subblocks
$\mathcal{C}^{(2)}_{1}$ and $\mathcal{C}^{(2)}_{2}$ contains two faulty level-$1$
qubits.

\subsection*{Level-2 decoding}

At level~$2$, each subblock $\mathcal{C}^{(2)}_{i}$ ($i=1,2$) therefore contains
errors on its physical qubits $4$ and $6$.
Since $4 \oplus 6 = 2$ for the Steane code, the local hard-decision decoder
applies a recovery on physical qubit~$2$, corresponding to the level-$1$
qubit $(i,2)_1$, thereby introducing a logical error on
$\mathcal{C}^{(2)}_{i}$.

Importantly, the physical cost of applying this logical flip on
$\mathcal{C}^{(1)}_{i,2}$ is $2$, because it cancels the existing recovery on the
level-$0$ qubit $(i,2,1)_0$.
Consequently, the total physical weight of the recovery on
$\mathcal{C}^{(2)}_{i}$ is $4$, which is smaller than the original error weight
$5$ on that subblock.
After level-$2$ decoding, both $\mathcal{C}^{(2)}_{1}$ and
$\mathcal{C}^{(2)}_{2}$ therefore carry nontrivial logical errors.

\subsection*{Level-3 decoding and baseline cost}

At level~$3$, the remaining errors occur on the level-$2$ qubits $(1)_2$ and
$(2)_2$.
Local hard-decision decoding therefore applies a correction on $(3)_2$, which
results in a logical error at the top level.

Within the bidirectional decoder, this corresponds to proposing a nontrivial
logical $X$ flip on the level-$2$ subblock $\mathcal{C}^{(2)}_{3}$.
The physical cost of realizing this level-$3$ logical flip is $9$.
Retaining the existing recoveries on $\mathcal{C}^{(2)}_{1}$ and
$\mathcal{C}^{(2)}_{2}$ (each of cost $4$), the total physical cost of this
baseline recovery is
\begin{equation}
	9 + 4 + 4 = 17.
\end{equation}

The bidirectional decoder next asks whether it is advantageous to transfer this
logical correction from $\mathcal{C}^{(2)}_{3}$ to the two subblocks
$\mathcal{C}^{(2)}_{1}$ and $\mathcal{C}^{(2)}_{2}$.

\subsection*{Cost estimation for reassignment}

We first consider applying a logical flip on $\mathcal{C}^{(2)}_{1}$.
Because there is an existing recovery on the level-$1$ qubit $(1,2)_1$, the
logical flip can be chosen to cancel this recovery, thereby reducing its
support from three level-$1$ qubits to two.
However, there are multiple stabilizer-equivalent realizations of such a
logical flip.

One possible realization supports the logical flip on the level-$1$ qubits
$(1,1)_1$, $(1,2)_1$, and $(1,3)_1$.
Since there are no existing recoveries on $\mathcal{C}^{(1)}_{1,1}$ and
$\mathcal{C}^{(1)}_{1,3}$, the physical cost of the corresponding logical flips
on these subblocks is $3$ each.
Combining these with the existing recoveries on
$\mathcal{C}^{(1)}_{1,2}$, $\mathcal{C}^{(1)}_{1,4}$, and
$\mathcal{C}^{(1)}_{1,6}$, the total physical cost of applying a logical flip on
$\mathcal{C}^{(2)}_{1}$ is
\begin{equation}
	3 + 3 + 1 + 1 + 1 = 9.
\end{equation}
An identical estimate applies to $\mathcal{C}^{(2)}_{2}$.
Thus, transferring the logical correction to both
$\mathcal{C}^{(2)}_{1}$ and $\mathcal{C}^{(2)}_{2}$ yields a total cost
\begin{equation}
	9 + 9 = 18 > 17,
\end{equation}
and the \textsf{FlipCost} subroutine causes the decoder to reject the
reassignment.

\subsection*{Alternative realization with lower physical cost}

The above conclusion depends on the particular stabilizer representative chosen
for the logical flips.
If instead the logical flip on $\mathcal{C}^{(2)}_{i}$ ($i=1,2$) is supported on
the level-$1$ qubits $(i,2)_1$, $(i,4)_1$, and $(i,6)_1$, it still cancels the
existing recovery on $(i,2)_1$.

In this case, applying logical flips on the level-$1$ subblocks
$\mathcal{C}^{(1)}_{i,4}$ and $\mathcal{C}^{(1)}_{i,6}$ has physical cost $2$
each, because the shifts cancel the existing recoveries on the level-$0$
qubits $(i,4,1)_0$ and $(i,6,1)_0$.
Including the remaining contribution, the total physical cost of the logical
flip on $\mathcal{C}^{(2)}_{i}$ is therefore
\begin{equation}
	2 + 2 + 1 = 5.
\end{equation}

Thus, transferring the logical correction to both
$\mathcal{C}^{(2)}_{1}$ and $\mathcal{C}^{(2)}_{2}$ admits a physical recovery of
total weight $10$, which exactly matches the original error pattern.
The failure of the bidirectional decoder in this example therefore originates
from the fact that \textsf{FlipCost} does not distinguish between stabilizer
choices that have identical intermediate-level weights but lead to different
physical costs, unless such choices are explicitly enumerated.

\end{document}